\documentclass[pra,reprint,twocolumn,showpacs]{revtex4-1}
\usepackage{color}
\usepackage{times}
\usepackage{amsmath}
\usepackage{amssymb}
\usepackage{amsfonts}
\usepackage{bm}
\usepackage{graphicx} 
\usepackage{multirow}

\usepackage[unicode,breaklinks]{hyperref}
\hypersetup{
    unicode=true,
    a4paper=true,
    plainpages=false,
    pdftitle={Numerical method for the stochastic projected Gross-Pitaevskii equation},
    pdfauthor={S. J. Rooney, P. B. Blakie, and A. S. Bradley},
    pdfsubject={Numerical method for the stochastic projected Gross-Pitaevskii equation},
    colorlinks=true,
    linkcolor=blue,
    citecolor=blue,
    filecolor=black,
    urlcolor=blue
}
\newcommand{\ve}{\varepsilon}
 \newcommand{\eref}[1]{(\ref{#1})} %short form
\newcommand{\eeref}[1]{Eq.~(\ref{#1})} %full form
\newcommand{\fref}[1]{Fig.~\ref{#1}}

\newcommand{\sref}[1]{section \ref{#1}}

\newcommand{\mint}[1]{\int d^3 #1\;}

\newcommand{\x}{\mathbf{x}}
\newcommand{\rr}{\mathbf{r}}

\newcommand{\kk}{\mathbf{k}}

\newcommand{\eq}[2]{\begin{equation}\label{#1}#2 \end{equation}}
\newcommand{\eqn}[1]{\begin{eqnarray}#1 \end{eqnarray}}

\newcommand{\EQ}[1]{\begin{eqnarray}#1\end{eqnarray}}
\newcommand{\EQsub}[1]{\begin{subequations}\begin{eqnarray}#1\end{eqnarray}\end{subequations}}

\newcommand{\rC}{\textbf{C}} % Classical region
\newcommand{\rI}{\textbf{I}} % Quantum region
\newcommand{\PC}{\mathcal{P}}

\newcommand{\LL}{\mathcal{L}}
\newcommand{\intV}[1]{\int d^3 #1\;}

\newcommand{\cf}{\psi} 
\def\x{\mathbf{r}}

\newcommand{\ecut}{\epsilon_{\rm cut}}

\definecolor{MyGray}{rgb}{.2,.5,.3}

\newcommand{\noise}[1]{\langle{#1}\rangle}

\begin{document}

\title{Numerical method for the  stochastic projected Gross-Pitaevskii equation}
\author{S. J. Rooney} 
\author{P. B. Blakie} 
\author{A.~S. Bradley} 
\address{Jack Dodd Centre for Quantum Technology, Department of Physics, University of Otago, Dunedin, New Zealand.}
\date{\today}
\begin{abstract}
We present a method for solving the stochastic projected Gross-Pitaevskii equation (SPGPE) for a three-dimensional Bose gas in a harmonic-oscillator trapping potential.  The SPGPE contains the challenge of both accurately evolving all modes in the low energy classical region of the system, and evaluating terms from the number-conserving scattering reservoir process.  We give an accurate and efficient procedure for evaluating the scattering terms using a Hermite-polynomial based spectral-Galerkin representation, which allows us to precisely implement the low energy mode restriction. Stochastic integration is performed using the weak semi-implicit Euler method.  We extensively characterize the accuracy of our method, finding a faster than expected rate of stochastic convergence. Physical consistency of the algorithm is demonstrated by considering thermalization of initially random states.  
\end{abstract}
\maketitle
%============================================================================
\section{Background}
\subsection{Introduction}
Providing a quantitative description of non-equilibrium dynamics of Bose-Einstein condensates (BECs) at finite temperature is an ongoing challenge \cite{Blakie08a,Proukakis08a}.  Classical field methods have been a popular tool to describe finite temperature BECs, utilizing the tractability of the Gross-Pitaevskii equation (GPE) to simulate the dynamics of many highly occupied modes of the system \cite{Davis2001b,Goral2002}.  These highly occupied \emph{coherent} modes are treated with a classical field approximation, enabling the dynamics of these modes to be treated nonperturbatively.  These methods have been used to treat of both equilibrium and non-equilibrium systems in a range of finite temperature systems \cite{Sinatra2001,Goral2002,Blakie05a,Davis2006a,Simula2006a,Wright08a,Bezett09a,Bezett09b,Bisset09c,Wright09a,Wright10a,Wright2011a}.   

Introducing coupling between the coherent region and the remaining \emph{incoherent} reservoir results in a stochastic GPE, which includes damping and noise terms from the reservoir interaction.  Such a description has been derived microscopically \cite{SGPEI,SGPEII,Stoof1997,Stoof1999}, allowing for these methods to provide an ab initio description of non-equilibrium dynamics.  There are now numerous examples in the literature of the application of stochastic GPEs to a range of systems, including vortex decay \cite{Rooney10a,Rooney11a,Rooney2012a}, soliton decay \cite{Cockburn10a,Cockburn11a}, defect formation across phase transitions \cite{Bradley08a,Weiler08a,Damski10a,Das2012a,Su2013a}, spinor condensates \cite{Su2012a,Liu2012a,Liu2012b,Song2013a}, polariton condensates \cite{Wouters09a}, equilibrium properties \cite{Garrett2013a}, and low dimensional systems \cite{Stoof2001,Duine2002,Proukakis06a,Proukakis06b,Cockburn09a,Cockburn11b,Cockburn11c,Gallucci12a,Cockburn12a,Davis12a}.  

Currently, most applications of stochastic GPEs have only included growth processes, where collisions between two incoherent region atoms lead to a change in population of the coherent region.  The damping term that arises from this process is similar to that of the damped GPE \cite{Choi1998,Penckwitt2002,Tsubota2002,Kasamatsu2003,Liu2013a}, and is relatively simple to implement numerically.  In the stochastic projected GPE (SPGPE) of Gardiner \emph{et al.} there are scattering reservoir processes, where a collision between coherent and incoherent atoms results in energy change with no population transfer.  These scattering terms have been recently implemented in Ref.~\cite{Rooney12a}, showing a dominant effect on highly non-equilibirum dynamics. 

In this paper, we present a numerical method for evolving the SPGPE, including the scattering terms.  Numerically solving the SPGPE involves two major technical challenges: (i) All moderately occupied coherent modes play an important role in finite temperature non-equilibrium dynamics.  Thus all modes beneath a well-chosen cutoff must be propagated accurately. (ii) The deterministic term arising from the reservoir interaction is non-local, while the noise is multiplicative and spatially correlated. Thus  accurately and efficiently implementing the scattering terms is a challenge.    Previous work has shown how a spectral approach \cite{Dion2003,Bao2005a} can be used within the PGPE formalism to precisely implement the energy cutoff,  and provide a means for accurate evolution of all low energy coherent modes for a Bose gas with both contact \cite{Blakie08b} and dipolar interactions \cite{Blakie2009a}.  In this work we extend these methods for the PGPE, providing a numerical method for evolving the full SPGPE that evaluates the scattering terms, while still propagating all coherent modes accurately.  We thus give a complete spectral-Galerkin method for the SPGPE.

This paper is organized as follows: In the rest of this section we briefly review the SPGPE formalism. In Sec.~\ref{sec:formalalgorithm} we outline our spectral approach, and outline the equations for the mode amplitudes we need to evaluate to solve the SPGPE.  In Sec.~\ref{matrixelements} we present our algorithm for evaluating the scattering terms using a Gauss-Hermite quadrature approach.  In Sec.~\ref{accuracy} we characterize the accuracy of our approach, with results demonstrating the convergence with quadrature grid size and evolution time step size.  We present an example of the usage of the SPGPE, showing the evolution of a random initial state to equilibrium.

\subsection{SPGPE theory}
We briefly outline the key formalism of the SPGPE relevant to our work (see Refs. \cite{SGPEI,SGPEII,Blakie08a,Rooney12a} for full details).  Here we consider a system of bosons confined in a three-dimensional trapping potential, described by the dimensionless single-particle Hamiltonian
\EQ{H_{\rm sp} &=& H_0 + \delta V(\x,t), }
where $\x = (x,y,z)$, and
\EQ{H_0 &=& -\frac{1}{2}\nabla^2 +\frac{1}{2} \sum_{j=1}^3 \lambda_j^2x_j^2,}
is the dominant contribution to the single particle Hamiltonian and defines our c-field basis, any time dependent perturbation potential is included in $\delta V(\x,t)$, and $\lambda_j = \omega_j/\omega_0$ is the relative trap frequency in each direction $j = \{x,y,z\}$.  Note that we work in dimensionless units throughout this paper, using harmonic oscillator units of length $x_0 = \sqrt{\hbar/m\omega_0}$, energy $E_0 = \hbar \omega_0$, and time $t_0 = 1/\omega_0$, where $m$ is the particle mass, and $\omega_0$ is a chosen reference frequency. 

In our c-field approach we divide the modes of our system into two subspaces according to their occupation, which we refer to as the \emph{coherent} (\rC) and \emph{incoherent} (\rI) regions \cite{Blakie05a}.   The low energy \rC~region contains highly occupied modes, which are dominated by classical fluctuations. In this case, we may use a classical field approximation to treat these highly occupied modes \cite{Blakie08a}. The remaining \rI~region contains sparsely occupied modes, and plays the role of a thermal reservoir.  The SPGPE is a stochastic equation of motion for the \rC-field, which accounts for the reservoir interactions from the \rI~region.  Treating the \rI~region semi-classically and assuming spatially constant reservoir interaction rates \cite{Bradley08a,Rooney12a}, the evolution equation is given in Stratonovich (S) form by \cite{Rooney12a}
\EQ{\label{SPGPE}
(S) d\psi(\rr,t)=d\psi\Big|_H+d\psi\Big|_\gamma+(S)d\psi\Big|_\ve,
}
with
\EQ{
\label{spgpeH}
d\psi\Big|_H&=&\PC\left\{-i\LL\psi dt\right\},\\
\label{spgpeG}
d\psi\Big|_\gamma&=&\PC\left\{\gamma(\mu-\LL)\psi dt+dW_\gamma(\x,t)\right\},\\
\label{spgpeE}
(S)d\psi\Big|_\ve&=&\PC\left\{-iV_\ve(\x,t)\psi dt+i\psi dW_\ve(\x,t)\right\},
}
where the evolution of the c-field is formally restricted to the \rC~region by the projector operator
\EQ{\PC f(\x) \equiv \sum_{n\in\rC} \phi_n(\x) \int d^3\x^\prime \phi^*_n(\x^\prime) f(\x^\prime), }
where $\phi_n(\x)$ are eigenstates of the single-particle Hamiltonian satisfying $ H_0 \phi_n =  \epsilon_n \phi_n$, and the summation includes all modes the \rC~region defined as
\eq{sumrestrict}
{\rC = \left\{ n:\epsilon_n \leq\epsilon_{\rm cut} \right\} ,}
where $\ecut$ is the single-particle energy cutoff defining the \rC~region. 

The first term of the SPGPE \eref{spgpeH} describes the Hamiltonian evolution, where 
\EQ{\LL\cf \equiv (H_{\rm sp}+ C_{\rm NL}|\cf|^2 )\cf,}
is the Hamiltonian evolution operator for the \rC~region, where the dimensionless nonlinearity constant is $C_{\rm NL} = 4\pi a/x_0$, where $a$ is the s-wave scattering length.  On its own \eeref{spgpeH} is the PGPE, a nonlinear Schr\"ondiger equation for the \rC-field as an isolated microcanonical system \cite{Blakie08a}.

The remaining terms in the SPGPE account for distinct reservoir interactions.

\subsubsection{Growth reservoir interaction}
\eeref{spgpeG} describes the growth reservoir interaction, where two \rI-region atoms collide leading to population growth of the \rC-region. The rate of this process is set by $\gamma$ \cite{Bradley08a}, and the Gaussian complex noise has the non-vanishing correlation
\EQ{\noise{dW^*_\gamma(\x,t) dW_\gamma(\x^\prime,t)} = 2 \gamma T \delta_\rC(\x,\x^\prime) dt, }
where 
\EQ{\delta_\rC(\x,\x^\prime) = \sum_{n\in\rC} \phi_n(\x) \phi_n^*(\x^\prime),\label{deltafuncC}}
is a delta function in the \rC~region. 

\subsubsection{Scattering reservoir interaction}
\eeref{spgpeE} describes the number conserving scattering reservoir interaction where energy and momentum are transferred between the \rC~and \rI~regions, without population transfer.  This process is described by the effective potential 
\eq{Ve}{ V_\ve(\x,t) = -\mathcal{M} \mint{\kk}\frac{e^{i\kk\cdot\x}}{(2\pi)^{3/2}} i\hat{\kk}\cdot\int d^3\x^\prime \frac{e^{i\kk\cdot\x^\prime}}{(2\pi)^{3/2}} \mathbf{j}(\x^\prime,t), }
where $\hat{\kk} = \kk/|\kk|$, the \rC-field current is 
\EQ{\label{current}\mathbf{j} (\x,t) = \frac{i}{2} \left( \cf \nabla \cf^* - \cf^* \nabla \cf \right),}
and 
\EQ{\label{Mdef}\mathcal{M} = \frac{16 \pi a^2}{x_0^2} \frac{1}{e^{\beta(\ecut - \mu)}-1} .}
The real scattering noise $dW_\ve$ has the non-zero correlation 
\EQ{\label{scattnoisecorr} \noise{dW_\ve(\kk,t) dW_\ve(\kk^\prime,t) } = \frac{2 \mathcal{M} T}{|\kk|} \delta_\rC (\kk,-\kk^\prime) dt,}
where the correlation is anti-diagonal in $k$-space so can be easily sampled numerically [see Sec.~\ref{sec:noise}].

\subsubsection{Scattering SPGPE}
By neglecting the growth terms in the SPGPE \eref{SPGPE}, we arrive at the scattering SPGPE
\EQ{\label{scattSPGPE} (S) d\cf &=& d\psi\Big|_H+(S)d\psi\Big|_\ve  \\
&=& -i \mathcal{P} \left\{ \left( \LL + V_\ve(\x,t) \right) \cf dt  - \cf dW_\ve (\x,t)  \right\} .}
In this work we are concerned with developing a method for evaluating the terms involved with the scattering processes.  Since including the growth terms is a relatively simple extension of the PGPE algorithm \cite{Bradley08a}, we will consider only the scattering SPGPE only from here onwards.

%%%%%%%%%%%%%%%%%%%%%%%%%%%%%%%%%%%%%%%%%%%%%%%%%%%
\section{Formal algorithm \label{sec:formalalgorithm}}

%%%%%%%%%%%%%%%%%%%%%%%%%%%%%%%%%%%%%%%%%%%%%%%%%%%
\subsection{Spectral representation}

We use the single-particle eigenstates  as our spectral basis, so we write the c-field as 
\eq{cfsumdmls}
{ \cf(\x,t)  = \sum_{n\in{\rC}} c_n(t) \phi_n( {\x}),}
where $c_n$ are time dependent complex amplitudes and $n$ represents all quantum numbers required to specify a single-particle state.  This choice is convenient because it allows us to efficiently implement the projection by restricting the spectral modes [as indicated in Eq.~(\ref{cfsumdmls})] to the set indicated in \eeref{sumrestrict} defining the \rC~region.

%%%%%%%%%%%%%%%%%%%%%%%%%%%%%%%%%%%%%%%%%%%%%%%%%%%
\subsection{Mode evolution}
\subsubsection{Spectral-Galerkin formulation}
We exploit that $H_0$ is diagonal in the spectral basis and use a Galerkin approach \cite{Dion2003}, where we project the scattering SPGPE \eref{scattSPGPE} onto the spectral basis.  This leads to a system of equations for the evolution of the amplitudes, i.e.
\begin{equation}
(S)\,\,dc_n=-i[\epsilon_nc_n+ G_n+S_n] dt +dB_n,\label{dcn} 
\end{equation}
where
\begin{align}
G_n&\equiv  C\int d^3\x\,\phi_n^*(\x)\,|\psi (\x,t)|^2\psi (\x,t),\label{Gn} \\
S_n&\equiv  \int d^3\x\,\phi_n^*(\x)V_{ {\ve}}(\x,t)\psi (\x,t), \label{Sn} \\
dB_{n}&\equiv -i\int d^3\x\,\left\{\phi_n^*(\x)\psi (\x,t) \sum_m \zeta_m(\x) dw_m\right\}\label{Bnm}, 
\end{align}
are the nonlinear  matrix elements of the two-body interaction, scattering effective potential, and scattering noise terms respectively.  We introduce the functions $\zeta_m(\x)$ later [see Sec.~\ref{sec:noise_matrixelements}], and  $dw_m$ is the standard real Wiener process satisfying  
\EQ{\label{dw1}\langle dw_n\rangle&=&0\\
\label{dw2}\langle dw_m dw_n\rangle&=&\delta_{mn}dt.}

There are two main steps in solving this equation: (i) time-evolution to step this equation forward in time [see Sec.~\ref{stochevo}]; and (ii) evaluating the non-linear matrix elements (\ref{Gn})-(\ref{Bnm}) at each time step [see Sec.~\ref{matrixelements}]. 

\subsubsection{Stochastic time evolution algortihm \label{stochevo}}
Because the noise associated with the scattering reservoir interaction is multiplicative, we use the \textit{weak vector semi-implicit Euler algorithm}   \cite{SM,Milstein,Drummond1991,Werner1997} to evolve our stochastic equations forward in time. As this algorithm is extensively discussed in the literature we briefly review the algorithm here. Equation (\ref{dcn}) is of the general form \begin{equation}
(S)\,dc_n=a_n(t,\mathbf{c})\,dt+ dB_{n}(t,\mathbf{c},dw_{m}),
\end{equation}
where $a_n=-i[\epsilon_n c_n+G_n+S_n]$, we use the notation $\mathbf{c}$ to represent the dependence of matrix elements on the full field ($\psi$), and the noise matrix elements depends on the Wiener process $dw_m$.  The solution is propagated to a set of discrete times $t_{j}=j\,\Delta t$, where $\Delta t$ is the step size, and we denote that solution at time  $t_j$ as $c^{(j)}_n$.  
Using this solution, the solution at the next time-step is computed as $c_n^{(j+1)}= c^{(j)}+\Delta c^{(j)}$, where
\begin{align}
 \Delta c^{(j)}_n&=a_n(\bar{t}_j,\bar{\mathbf{c}}^{(j)})\,\Delta t +dB_{n}(\bar{t}_j,\bar{\mathbf{c}}^{(j)},\Delta w_m^{(j)}),\label{SIE}
\end{align}
with 
\begin{align}
\bar{c}^{(j)}_n&\equiv\frac{1}{2}(c^{(j)}_n+c^{(j+1)}_n),\label{cbar}\\
\bar{t}_j&\equiv\frac{1}{2}(t_{j+1}+t_{j}),  \\
\langle \Delta w^{(j)}_m\Delta w^{(j)}_n\rangle &=\Delta t\, \delta_{mn}.\label{Deltaw}
\end{align}
Note formally $\Delta w^{(j)}_n\equiv\int_{t_j}^{t_{j+1}}dw_n$, however in practice we sample $\Delta w^{(j)}_m$ as a real Gaussian distributed random variable of variance $\Delta t$ [c.f.~Eq.~(\ref{Deltaw})].

%%%%%%%%%%%%%%%%%%%%%%%%%%%%%%%%%%%%%%%%%%%%%%%%%%%
\section{Evaluation of the scattering SPGPE matrix elements \label{matrixelements}}
Here we give a full description of our algorithm to efficiently and accurately evaluate the scattering SPGPE matrix elements in the harmonic oscillator basis.  This expands on the brief overview of our method presented in Ref.~\cite{Rooney12a}.
 
\subsection{Harmonic-oscillator state properties}
We firstly discuss some important properties of our single-particle basis used in this work.  

\subsubsection{Seperability}
The eigenstates of the basis Hamiltonian ($H_0$) are separable into one dimensional basis states, that is,
\eqn{
 \phi_n(\x) &\leftrightarrow& \phi^{\lambda_x}_\alpha( x) \phi^{\lambda_y}_\beta( y) \phi^{\lambda_z}_\gamma( z), \\
\epsilon_n &\leftrightarrow& \varepsilon_\alpha + \varepsilon_\beta + \varepsilon_\gamma, \\
c_n &\leftrightarrow& c_{\alpha \beta \gamma},}
where $\phi^{\lambda_x}_\alpha(x)$ are eigenstates of the dimensionless 1D harmonic-oscillator Hamiltonian
\eq{HO1D}
{\left[ -\frac{1}{2} \frac{d^2}{d  x^2} + \frac{1}{2} \lambda_x^2 x^2 \right]  \phi^{\lambda_x}_\alpha( x) =  \varepsilon_\alpha  \phi^{\lambda_x}_\alpha( x), }
where the harmonic-oscillator states take the form
\eq{basisstates}
{\phi^{\lambda_x}_\alpha ( x) = \lambda_x^{1/4} h_\alpha H_\alpha(\sqrt{\lambda_x} x) e^{- \lambda_x x^2/2} ,}
where the normalization constant is $h_\alpha = [2^\alpha \alpha! \sqrt{\pi}]^{-1/2}$, and $H_\alpha ( x)$ is a Hermite polynomial of degree $\alpha$, defined by the recurrence relation
\eq{Hermite}
{H_{\alpha+1} ( x) = 2 x H_\alpha ( x) - 2 \alpha H_{\alpha - 1} ( x), \;\; \alpha = 1,2,...}
with $H_0 ( x) = 1,$ and $H_1( x) = 2 x$.  The eigenvalue is the single-particle energy given by $ \epsilon_\alpha = \lambda_x (\alpha + \frac{1}{2})$, where $\alpha$ is a non-negative integer (we use Greek subscripts to denote 1D eigenstates). The \rC~region is defined by the region containing all modes below the single-particle energy cutoff
\eq{Cregion1D}
{\rC = \{ \alpha,\beta,\gamma :  \varepsilon_\alpha +  \varepsilon_\beta +  \varepsilon_\gamma \leq  \epsilon_{\rm cut} \} ,}
so that within the \rC~region there exist $M_x ~(\approx  \epsilon_{\rm cut})$ distinct 1D eigenstates in each direction, and $M_T \approx M_x M_y M_z /6$ total 3D basis states in the \rC~region. 

For this work we will consider a spherically symmetric system ($\lambda_x = \lambda_y = \lambda_z = 1$), to avoid cumbersome notation involving $\lambda$.  However the work presented in this paper can be easily generalized for any trap anisotropy, by retaining the general form of \eeref{basisstates}.  Thus we drop any reference to $\lambda$, and consider a system with $M$ modes in the \rC~region in each direction.

\subsubsection{Step operators \label{stepops}}
Step operators allow us to represent certain operators \emph{exactly} in the spectral basis.  The step operators are defined as
\EQ{\hat{a}^+_x &=& \frac{1}{\sqrt{2}} \left(  -\frac{\partial}{\partial x} + x \right), \\
\hat{a}^-_x  &=&  \frac{1}{\sqrt{2}} \left(  \frac{\partial}{\partial x} + x \right) . }
We then find that the matrix representation of the step operators in the spectral basis is
\EQ{ (\hat{a}_x^+ )_{\alpha\beta} &\equiv& \int dx \phi_\alpha^*(x) \hat{a}^+_x  \phi_\beta (x) \\
&=& \sqrt{\beta+1} \delta_{\alpha,\beta+1},}
and similarly,
\EQ{\hat{a}_x^+= \sqrt{\beta} \delta_{\alpha,\beta-1}.}
For our purposes, this allows us to differentiate in the spectral basis exactly, since
\EQ{(\hat\partial_x)_{\alpha \beta} &=& \frac{1}{\sqrt{2}} (a_x^- - a_x^+)_{\alpha\beta}\\
&=& \sqrt{\frac{\beta}{2}} \delta_{\alpha,\beta-1} - \sqrt{\frac{\beta+1}{2}} \delta_{\alpha,\beta+1}.\label{eq:diff_basis}}
This 1D procedure is applied in the same manner to include $y$ and $z$. It is important that we can apply the derivative operator exactly in the spectral basis, since we need to evaluate the c-field current \eref{current} to evaluate the scattering effective potential term. 

For a further discussion on the use of step operators to calculate observables, see Ref.~\cite{Blakie08b}.

%%%%%%%%%%%%%%%%%%%%%%%%%%%%%%%%%%%%%%%%%%%%%%%%%%%
\subsection{Two-body interaction term \label{interactionterm}}
Here we briefly review our scheme for numerically integrating the nonlinear contact interaction term (\ref{Gn}) using Gauss-Hermite quadrature, which is described in complete detail in Ref.~\cite{Blakie08b}.  This algorithm calculates the nonlinear interaction term exactly for a harmonically trapped system, and is required to integrate the scattering SPGPE \eref{scattSPGPE}.  

Firstly, using (\ref{cfsumdmls}) and the form of the basis states (\ref{basisstates}), we can write the c-field as 
\eq{cfpoly}
{\cf = Q( \x) e^{-(x^2+y^2+z^2)^2/2},}
where
\eq{Qpoly}
{Q( \x) \equiv\sum_{\{ \alpha \beta \gamma\}\in \rC } c_{\alpha \beta \gamma}( t) h_\alpha H_\alpha( x) h_\beta H_\beta( y) h_\gamma H_\gamma ( z) }
is a polynomial of maximum degree $M - 1$ in each independent coordinate due to the energy cutoff.  

The interaction term (\ref{Gn}) is forth order in the field, so we can write it in the form
\eq{Gnpoly}
{G_{\alpha \beta\gamma} = \intV{ \x} e^{-2( x^2 + y^2 +  z^2)} P_{\alpha\beta\gamma} ( x, y, z),}
where
\eqn{\label{Ppoly}
P_{\alpha\beta\gamma} ( x,  y, z) &\equiv& C_{\rm NL}  h_\alpha H_\alpha( x) h_\beta H_\beta( y) h_\gamma H_\gamma ( z) \nonumber \\
&&\times | Q( x, y, z) |^2 Q( x, y, z) }
is a polynomial of maximum degree $4(M_x-1)$ in each coordinate.  We evaluate (\ref{Gnpoly}) using Gauss-Hermite quadrature.  The general form of the $N_Q$ point quadrature rule is
\eq{GaussHermQuad}
{\int^\infty_{-\infty} d x W( x) f( x) \approx \sum_{j=1}^{N_Q} w_j f( x_j), }
where $W( x)$ is a Gaussian weight function, and $w_j$ and $x_j$ are the quadrature weights and roots.  The Gauss-Hermite quadrature is exact if $f(x)$ is a polynomial of maximum degree $2N_Q-1$. Since the exponential in (\ref{Gnpoly}) takes the form of the appropriate weight function for Gauss-Hermite quadrature, we can evaluate the interaction matrix element \emph{exactly} by
\eq{Gnquad}
{G_{\alpha\beta\gamma} = \sum_{ijk}w_iw_jw_kP_{\alpha\beta\gamma}( x_i, x_j, x_k),}
using a three-dimensional spatial grid with $2(M-1)$ points in each direction, where the $\{x_i\}$ and $\{w_i\}$ are the $2(M-1)$ roots and weights of the one-dimensional Gauss-Hermite quadrature with weight function $W( x) = e^{-2 x^2}$.

%%%%%%%%%%%%%%%%%%%%%%%%%%%%%%%%%%%%%%%%%%%%%%%%%%%
\subsection{Scattering effective potential term \label{sec:algorithm_scattpot}}
To compute the effective potential matrix elements (\ref{Bnm}), we adapt the scheme developed for evaluating the dipolar interaction term in the PGPE~\cite{Blakie2009a}.    Here the calculation involves the Fourier transform of the current (\ref{current}), whose momentum space form is then Fourier transformed to form the effective potential \eref{Ve}.  As we are using a nonuniform quadrature grid to represent the c-field, we follow Ref.~\cite{Blakie2009a} in using an auxiliary harmonic-oscillator basis to perform the Fourier transforms.

Firstly we  calculate the c-field current density (\ref{current}) by
\EQ{\mathbf{j}(\x) = \sum^3_{v=1} j_v(\x) {\bf e}_v  \label{currentsum} }
where the index $v = \{x,y,z\}$ represents each spatial dimension with corresponding unit vector $ {\bf e}_v$, and
\EQ{\label{current_component} j_v(\x) = \frac{i}{2}\left( \cf \partial_v \cf^* - \cf^* \partial_v \cf \right)  ,}
where \eeref{current_component} can be calculated exactly using step operators [see \sref{stepops}]. Now using similar arguments to section \ref{interactionterm}, each component of the current can be written in the form
\eq{Jpoly}
{ j_v (\x) = R_v( x, y, z) e^{-( x^2 +  y^2 +  z^2)} ,}
where $R_i$ is polynomial of maximum degree $2(M-1)$ in each coordinate (due to the projector).

Following Ref. \cite{Blakie2009a}, we introduce a set of auxiliary harmonic-oscillator states,
\eq{auxHO}
{\chi_\alpha ( x) = \bar h_\alpha \bar H _\alpha ( x) e^{- x^2},}
where the exponential argument is chosen to match Eq.~(\ref{Jpoly}). These states are eigenstates of the harmonic oscillator Hamiltonian, with a trapping potential twice as tight as that defining the spectral basis (\ref{HO1D}).  Noting this, the auxiliary states are related to the spectral basis modes by $\chi_\alpha(  x) = 2^{1/4} \phi_\alpha (\sqrt{2}  x)$.

Due to the same exponential factor, we can represent each component of the current [i.e. \eref{current_component}] as
\EQ{j_v(\x) = \sum_{\alpha\beta\gamma} d^v_{\alpha\beta\gamma} \chi_\alpha ( x) \chi_\beta( y) \chi_\gamma( z),}
where $d^v_{\alpha\beta\gamma}$ is a set of $8M^3$ auxiliary basis coefficients for each component of $\mathbf{j}(\x)$.  Since the auxiliary oscillator states form an orthonormal basis, we can evaluate these basis coefficients by
\eqn{d^v_{\alpha\beta\gamma} &=& \intV{\x} \chi^*_\alpha ( x)\chi^*_\beta ( y) \chi^*_\gamma( z) j_v(\x) ,  \\
&=&\intV{\x} e^{-2( x^2 + y^2 +  z^2)} Z^v_{\alpha\beta\gamma} ( x, y, z)  ,      \label{dspectral}}
where
\eq{Spoly}
{Z^v_{\alpha\beta\gamma} ( x, y, z) = e^{2( x^2 + y^2 +  z^2)} \chi^*_\alpha ( x)\chi^*_\beta ( y) \chi^*_\gamma( z) j_v(\x) , }
is a polynomial of maximum degree $4(M-1)$ in each spatial dimension.  We can use the same Gauss-Hermite quadrature to integrate (\ref{dspectral}) as we used for the nonlinear interaction term (\ref{Gnpoly}), as the weight function and maximum polynomial degree are both $W(x) = e^{-2x^2}$.  Thus Eq.~(\ref{dspectral}) can be calculated exactly for each component of the current, by
\eq{dquad}
{d^v_{\alpha\beta\gamma} = \sum_{ijk} w_i w_j w_k Z^v_{\alpha\beta\gamma} ( x_i, y_j, z_k) , }
where the quadrature roots $\{x_i\}$, and weights $\{w_i\}$, are the same as used in (\ref{Gnquad}).

To evaluate the effective potential (\ref{Ve}), we must Fourier transform each component of the current.  Since the harmonic-oscillator states are eigenstates of the Fourier transform operator, 
\eq{FTop}
{\chi_\alpha ( k_x) = (-i)^{-\alpha} \frac{1}{(2\pi)^{1/2}} \int d x e^{-i  k_x  x} \chi_\alpha ( x),}
we can efficiently evaluate the Fourier transform with knowledge of auxiliary basis amplitudes of the current $(d^v_{\alpha\beta\gamma})$ by
\eq{jknumeric}
{ \tilde{j}_v ( \kk) = \sum_{\alpha\beta\gamma} (-i)^{-(\alpha+\beta+\gamma)} d^v_{\alpha\beta\gamma} \chi_\alpha ( k_x) \chi_\beta ( k_y)\chi_\gamma( k_z).}
Thus (\ref{jknumeric}) will represent the Fourier transform of the current on the quadrature grid exactly.  We can now evaluate the scattering effective potential term (\ref{Ve}) by evaluating
\EQ{\label{jkint}\Phi(\kk) =\hat{\kk} \cdot {\bf j}(\kk) = \frac{k_x \tilde{j}_x (\kk)+k_y\tilde{j}_y (\kk)+k_z\tilde{j}_z (\kk)}{|\kk|} ,}
and computing the inverse Fourier transform of \eeref{jkint}. To compute the inverse Fourier transform, we expand $ V_{\ve} (\x)$ in the auxiliary basis
\eq{VMaux}
{ V_\ve (\x) \approx \sum_{\alpha\beta\gamma} f_{\alpha\beta\gamma} \chi_\alpha ( x) \chi_\beta( y) \chi_\gamma( z),}
where 
\eq{fbasis}
{f_{\alpha\beta\gamma} = -i \mathcal{M} \intV{ \kk} (i)^{-(\alpha+\beta+\gamma)}\chi^*_\alpha ( k_x) \chi^*_\beta ( k_y)\chi^*_\gamma( k_z) \Phi(\kk) .}
We integrate (\ref{fbasis}) using a Gauss-Hermite quadrature, via
\eq{fquad}
{f_{\alpha\beta\gamma} = -i \mathcal{M}  \sum_{ijk}   \bar w_i  \bar w_j  \bar w_k  T_{\alpha\beta\gamma} (   k_i,   k_j,   k_k) ,}
where 
\eqn{\label{Tquad}
T_{\alpha\beta\gamma} (   k_i,   k_j,   k_k) &=& e^{2(   k_i^2+   k_j^2+   k_k^2)} \chi^*_\alpha (   k_i)\chi^*_\beta (   k_j) \chi^*_\gamma( \bar k_k) \nonumber \\ 
&&\times \Phi(  k_i,  k_j,  k_k). }
Here the quadrature roots $\{k_i\}$ and weights $\{\bar w_i\}$ are found from the weight function $e^{-2 k_i^2}$. Since in general $\Phi(\kk)$ cannot be represented exactly in the oscillator basis, our quadrature rule (\ref{fbasis}), and thus \eeref{VMaux}, are approximations.  Hence the number of $k$-grid quadrature points required is somewhat arbitrary [see Sec.~\ref{kspace_grids}].  We investigate how the accuracy of the scattering matrix element depends on the number of quadrature points in Sec.~\ref{accuracy}.

From the form of \eeref{VMaux}, we see that the effective potential matrix elements we require \eref{Sn} are of the form
\EQ{\label{Sint}S_{\alpha \beta \gamma}=  \intV{ \x} e^{-2( x^2 + y^2 +  z^2)} Y_{\alpha\beta\gamma} ( x, y, z), }
where 
\EQ{Y_{\alpha\beta\gamma} (\x) = h_\alpha H_\alpha(x)h_\beta H_\beta (y) h_\gamma H_\gamma(z) V_\ve(\x) Q(\x),}
where $Q(\x)$ represents the \emph{c}-field and is given by \eeref{Qpoly}.  Although our expression for $V_\ve(\x)$ is approximate, once in this form, $Y_{\alpha \beta \gamma}$ takes the form of a polynomial of degree $4(M - 1)$ in each coordinate.  Thus we can integrate \eeref{Sint} exactly using the same Gauss-Hermite quadrature sum as for \eeref{Gnquad}, i.e. using a three-dimensional spatial grid with $2(M-1)$ points in each direction, where the $\{x_i\}$ and $\{w_i\}$ are the $2(M-1)$ roots and weights corresponding to the weight function $W(x) = e^{-2 x^2}$.

%%%%%%%%%%%%%%%%%%%%%%%%%%%%%%%%%%%%%%%%%%%%%%%%%%%
\subsection{Scattering noise term \label{sec:noise}}
\subsubsection{Correlation function}
A noise with the necessary correlations \eref{scattnoisecorr} can be constructed as
\eq{noiseExp}{
d\tilde{W}_\ve(\kk)=\sqrt{\frac{2{\cal M} {T}dt}{|\kk|}}\sum_{\alpha\beta\gamma}dw_{\alpha\beta\gamma}\tilde{\phi}_{\alpha\beta\gamma}(\kk).
}
Here we have made use of the Fourier transformed modes $\tilde{\phi}_{\alpha\beta\gamma}(\kk)=(-i)^{(\alpha+\beta+\gamma)}\phi_{\alpha\beta\gamma}(\kk)$.  The phase factors arising in Fourier transforming the modes generates the antidiagonal delta-function we require, since 
\eqn{
\sum_{\alpha\beta\gamma}(-1)^{(\alpha\beta\gamma)}\phi_{\alpha\beta\gamma}(\kk)\phi_{\alpha\beta\gamma}(\kk)&=&\sum_{\alpha\beta\gamma}\phi_{\alpha\beta\gamma}(\kk)\phi_{\alpha\beta\gamma}(-\kk) \;\;\; \nonumber\\
&=&\delta_C(\kk,-\kk),\label{antid}}
where we have used \eeref{deltafuncC}.

\subsubsection{Computing the matrix elements \label{sec:noise_matrixelements}}
To implement the noise in our Galerkin approach, we need to evaluate the matrix elements corresponding to \eeref{Bnm}. This means we need to project the noise in position space onto our spectral basis.  Since numerically we form the correct scattering noise correlation in Fourier space [see \eeref{noiseExp}], we  need to compute
\EQ{dW_\ve (\x,t) &=& \mint{\kk}\frac{e^{i\kk\cdot\x}}{(2\pi)^{3/2}}d\tilde{W}_\ve(\kk) ,  \\
&=&\sum_{\alpha\beta\gamma} \zeta_{\alpha\beta\gamma} (\x) dw_{\alpha\beta\gamma},}
where the $\{dw_{\alpha\beta\gamma}\}$ are real Gaussian distributed random variables [see \eeref{dw1}-\eref{dw2}], and where
\eq{zetanoise}{\zeta_{\alpha\beta\gamma}(\x) \equiv \mint{\kk}\frac{e^{i\kk\cdot\x}}{(2\pi)^{3/2}} \Theta_{\alpha\beta\gamma} (\kk),}
where
\EQ{\Theta_{\alpha\beta\gamma}(\kk)=\sqrt{\frac{2\mathcal{M}T}{|\kk|}}\tilde{\phi}_\alpha(k_x)\tilde{\phi}_\beta(k_y)\tilde{\phi}_\gamma(k_z) , \label{theta_noise}}
where the $\{\tilde{\phi}_\alpha(k_x)\}$ are the Fourier transformed basis states.  Thus to evaluate the scattering noise matrix elements \eref{Bnm}, we must apply $\zeta_{\alpha\beta\gamma}$ to the Gaussian distributed random variables.  We now outline how we evaluate the scattering noise matrix elements.

Firstly, we can represent \eeref{theta_noise} in the form 
\EQ{\Theta_{\alpha\beta\gamma} (\kk) = X_{\alpha\beta\gamma}(\kk) e^{-(k_x+k_y+k_z)/2} , \label{dWe_spec} }
where 
\EQ{X_{\alpha\beta\gamma}(\kk) &=& (i)^{-(\alpha+\beta+\gamma)} h_\alpha  H_\alpha (k_x) h_\beta  H_\beta (k_y)    \nonumber\\ 
&&\times h_\gamma  H_\gamma (k_z)\sqrt{\frac{2\mathcal{M}T}{|\kk|}} . \label{noise_Hermites}}
In order to compute the inverse Fourier transform of \eeref{dWe_spec} to form $\zeta$, we use a similar process as in Sec.~\ref{sec:algorithm_scattpot}.    Here we do not need to use the auxiliary basis states necessary for computing $\tilde{\mathbf{j}}(\kk)$, since the exponential term in \eeref{dWe_spec} matches that of the standard basis states.  Thus, we can calculate $\zeta(\x)$ using the expansion of $\zeta(\x)$ in the standard spectral basis
\eq{zeta_sum}{\zeta_{\alpha\beta\gamma} (\x) \approx n_{\alpha\beta\gamma} \phi_\alpha (x) \phi_\beta (y) \phi_\gamma (z),  }
where
\eq{noise_basis_int}{n_{\alpha\beta\gamma} = \mint{\kk}(i)^{-(\alpha+\beta+\gamma)} \phi_\alpha(k_x) \phi_\beta(k_y) \phi_\gamma(k_z)   \Theta_{\alpha\beta\gamma} (\kk).} 
\eeref{zeta_sum} is approximate [c.f. \eeref{VMaux}] since we can not represent $X_{\alpha\beta\gamma}(\kk)$ as a polynomial.   We evaluate the integral using a Gauss-Hermite quadrature 

\EQ{n_{\alpha\beta\gamma} = \sum_{ijk} \tilde{w}_i \tilde{w}_j \tilde{w}_k U_{\alpha\beta\gamma} (\tilde k_i,\tilde k_j,\tilde k_k),}
where 
\eqn{\label{Upoly}U_{\alpha\beta\gamma} (\tilde k_i,\tilde k_j,\tilde k_k) &=& e^{( \tilde k_i^2+ \tilde k_j^2+ \tilde k_k^2)} \phi^*_\alpha ( \tilde k_i)\phi^*_\beta ( \tilde k_j) \phi^*_\gamma( \tilde k_k)\nonumber\\
&&\times\Theta(\tilde k_i,\tilde k_j,\tilde k_k) ,}
where the quadrature roots $\{\tilde k_i\}$ and weights $\{\tilde{w}_i\}$ correspond to the weight function $W = e^{-k_i^2}$.  Since our Gauss-Hermite quadrature is approximate, the number of $k$ points is again somewhat arbitrary here.  See Sec.~\ref{accuracy} for further discussion.

 Finally, since the matrix elements we require \eref{Bnm} are third order in the field, they can be cast in the form
 \eq{Bnmintegral}{dB_{\alpha\beta\gamma} = \mint{\x} e^{-\frac{3}{2}(x^2+y^2+z^2)} U_{\alpha\beta\gamma},  }
 where 
\eqn{\label{Upoly}U_{\alpha\beta\gamma}(\x) &=& h_\alpha H_\alpha(x) h_\beta H_\beta(y) h_\gamma H_\gamma(z)   \nonumber\\
&&\times Q(\x) \sum_{\kappa\lambda\mu}\zeta_{\kappa\lambda\mu}(\x) dw_{\kappa\lambda\mu},}
where $Q(\x)$ represents the c-field and is given by \eeref{Qpoly}, so $U_{\alpha\beta\gamma}(\x)$ is a polynomial of degree $3(M_x-1)$ in each coordinate.  Thus we can evaluate \eeref{Bnmintegral} using a Gauss-Hermite quadrature 
\eq{Bnmquad}{B_{\alpha\beta\gamma}= \sum_{ijk} \hat{w}_i \hat{w}_j \hat{w}_k U_{\alpha\beta\gamma}(\hat{x}_i,\hat{x}_j,\hat{x}_k) .}
This quadrature is exact using a three-dimensional grid with $3(M_x-1)/2$ quadrature points in each coordinate, where the grid points $\{\hat{x}_i\}$ and quadrature weights $\{\hat{w}_i \}$ correspond to the weight function $W(x) = e^{-3x^2/2}$.

%%%%%%%%%%%%%%%%%%%%%%%%%%%%%%%%%%%%%%%%%%%%%%%%%%%
\subsection{Summary of the algorithm \label{sec:algorithm}}
Here we give a summary of our algorithm for calculating the scattering SPGPE matrix elements.  We split our summary into two sections, first dealing with the deterministic terms followed by the noise term. 

\subsubsection{Fourth order field matrix elements - $G_n$ and $S_n$ \label{sec:scattpotoveriew}}

For each Euler step, the procedure for calculating these matrix elements is as follows:
\begin{itemize}
\item[Step 1:] We firstly transform the c-field from the spectral to spatial representation via
\eqn{\cf(\x_{\bm{s}}) = \sum_{\bm{\sigma}} U_{\bm{s \sigma}} c_{\bm{\sigma}}, \label{eq:basis2pos}}
where $\bm{\sigma} = \{\alpha\beta\gamma\}$,  $\x_{\bm{s}} = (x_i,x_j,x_k)$ are the quadrature points associated with the weight function $W(x) = e^{-2x^2}$, and $U_{\bm{s \sigma}} = U_{i\alpha} U_{j\beta} U_{k\gamma}$ are the precomputed transformation matrices.  These are the one-dimensional basis states evaluated at the quadrature grid points
\eqn{U_{i\alpha}=\phi_\alpha(x_i) .}

\item[Step 2:] Differentiate the field
 \EQsub{ c^\prime_{x\bm{\sigma}} &=&\sum_{\kappa\lambda\mu} (\hat{\partial}_x)_{\alpha\kappa} (\hat{I}_y)_{\beta\lambda} (\hat{I}_z)_{\gamma\mu} c_{\kappa\lambda\mu}, \\
c^\prime_{y\bm{\sigma}} &=&\sum_{\kappa\lambda\mu} (\hat{I}_x)_{\alpha\kappa} (\hat{\partial}_y)_{\beta\lambda} (\hat{I}_z)_{\gamma\mu} c_{\kappa\lambda\mu}, \\
c^\prime_{z\bm{\sigma}} &=&\sum_{\kappa\lambda\mu} (\hat{I}_x)_{\alpha\kappa} (\hat{I}_y)_{\beta\lambda} (\hat{\partial}_z)_{\gamma\mu} c_{\kappa\lambda\mu}, }
where the derivative operators are given by \eeref{eq:diff_basis}, and the $\hat{I}$ are the identity operators. 
\item[Step 3:] Transform the differentiated fields to position space
 \EQsub{ \cf^\prime_x(\x_{\bm{s}}) &=&\sum_{\bm{\sigma}} U_{\bm{s \sigma}}c^\prime_{x\bm{\sigma}}, \\
\cf^\prime_y(\x_{\bm{s}}) &=& \sum_{\bm{\sigma}} U_{\bm{s \sigma}}c^\prime_{y\bm{\sigma}}, \\
\cf^\prime_z(\x_{\bm{s}}) &=& \sum_{\bm{\sigma}} U_{\bm{s \sigma}}c^\prime_{z\bm{\sigma}}, }
so we can then form the position space current current via \eeref{currentsum}, where we have 
\EQ{\label{current_component_quadgrid} j_v(\x_{\bm{s}}) = \frac{i}{2}\left[ \cf(\x_{\bm{s}}) \cf_v^\prime(\x_{\bm{s}})^* - \cf(\x_{\bm{s}})^*  \cf_v^\prime(\x_{\bm{s}}) \right] .}

\item[Step 4:] The weighted position space current is constructed for each component of $\mathbf{j}(\x_{\bm{s}})$
\begin{subequations}
\EQ{{J}_x( \x_{\bm{s}}) &\equiv& w_{\bm{s}} e^{2\x_{\bm{s}}^2}  j_x (  \x_{\bm{s}}), \\
{J}_y(\x_{\bm{s}}) &\equiv& w_{\bm{s}}  e^{2\x_{\bm{s}}^2}  j_y (  \x_{\bm{s}}),\\
{J}_z( \x_{\bm{s}}) &\equiv& w_{\bm{s}}  e^{2\x_{\bm{s}}^2}  j_z (  \x_{\bm{s}}),}
\end{subequations}
where $w_{\bm{s}} = (w_i,w_j,w_k)$ are the quadrature weights associated with the weight function $W(x) = e^{-2x^2}$.

\item[Step 5:]  We compute the Fourier transform of each component of the current by
\EQsub{ \tilde{j}_{k_x} (  \kk_{\bm{t}})&=& \sum_{\bm{s}} W_{\bm{st}}  {J}_x( \x_{\bm{s}}), \\ 
\tilde{j}_{k_y} (  \kk_{\bm{t}})&=& \sum_{\bm{s}}  W_{\bm{st}}  {J}_y(  \x_{\bm{s}}), \\ 
\tilde{j}_{k_z} (  \kk_{\bm{t}})&=& \sum_{\bm{s}} W_{\bm{st}}  {J}_z( \x_{\bm{s}}). }
Here the precomputed transformation matrix is
\eq{Wtransf}
{W_{ir} = \sum_{\alpha} (-i)^\alpha \chi_\alpha (   k_r) \chi_\alpha (  x_i) ,}
where the $k$-space quadrature grid is based on the weight function $W(k) = e^{-2k^2}$. \eeref{Wtransf} combines both steps of the Fourier transform into one operation [$  j_x(\x) \to d_{\alpha\beta\gamma}$, and $d_{\alpha\beta\gamma}\to   \tilde{j}_{kx}(\kk)$].  

\item[Step 6:] In Fourier space we form the projection of the current onto $\hat{ {\kk}}$.  We then premultiply with the quadrature weights to form
\eqn{ \tilde{\mathcal{F}}(\kk_{\bm t}) \equiv \tilde w_{\bm t}  e^{2\kk_{\bm t}^2}  \hat{  \kk}_{\bm t} \cdot  \tilde{\mathbf j} (  \kk_{\bm t}). }

\item[Step 7:] Inverse transforming takes us back to position space. and computes the effective potential $V_\ve$:
\eq{Phielement}
{V_\ve (\x_{\bm s}) = -i \mathcal{M} \sum_{\bm t} W_{\bm{st}}^* \tilde{\mathcal{F}}(\kk_{\bm t}).}

\item[Step 8:] We combine the two-body interaction and scattering effective potential terms into a single integrand (since the quadrature sum has the same weight function for both cases),
\eqn{g(\x_{\bm s}) = w_{\bm s} e^{2\x_{\bm s}^2} \left[C_{\rm NL} |\psi(\x_{\bm s})  |^2 + V_\ve (\x_{\bm s})  \right]\psi(\x_{\bm s})   \label{fintegrand} .} 

\item[Step 9:] Inverse transforming $f(\x_k)$ completes the integration of \eeref{Gn} and \eeref{Sn}, giving the required matrix elements
\eq{Fcompute}
{G_{\bm \sigma} + S_{\bm \sigma} = \sum_{\bm s} U^*_{\bm{s \sigma}}  g(\x_{\bm s}).}

\end{itemize}

%%%%%%%%%%%%%%%%%%%%%%%%%%%%%%%%%%%%%%%%%%%%%%%%%%%
\subsubsection{Third order field matrix elements - Scattering noise $dB_{n}$ \label{sec:scattnoiseoveriew}}
To implement the noise, we generate $dw_{\bm{\sigma}}$ at the start of each Euler step.  Then for each iteration of \eeref{SIE}, we perform the following steps:
\begin{itemize}
\item[Step 1:] We transform the random variables to Fourier space via
\EQ{\Psi(\tilde\kk_{\bm{t}}) = \sum_{\bm{\sigma}} \tilde{U}_{\bm{t\sigma}} dw_{\bm{\sigma}} }
where the precomputed transformation matrices are the Fourier transformed basis states 
\EQ{\tilde{U}_{\bm{t\sigma}} = (-i)^{(\alpha+\beta+\gamma)} \phi_\alpha(\tilde k_u)  \phi_\beta (\tilde k_v) \phi_\gamma (\tilde k_w), }
where the tilde notation implies the $k$-space quadrature grid is based on the weight function $W(k) = e^{-k^2}$.

\item[Step 2:] We form the appropriate correlation function in $k$-space, and then compute the inverse Fourier transform:
\EQ{f(\hat\x_{\bm{s}}) = \sum_{\bm{t}} X^*_{\bm{st}}  \sqrt{ \frac{2 T \mathcal{M}}{|\tilde\kk_{\bm{t}}|}}  \Psi(\tilde\kk_{\bm{t}}) \label{fnoise} } 
where 
\EQ{X_{ir} = \sum_{\alpha} (-i)^\alpha \phi_\alpha (\tilde k_r) \phi_\alpha (\hat{x}_i),}	
where the hat notation implies the position space quadrature grid is based on the weight function $W(x) = e^{-3x^2/2}$.

\item[Step 3:] We now need to transform the \emph{c}-field to position space, onto the grid based on the three-field weight function:
\EQ{\cf(\hat\x_{\bm s})  = \sum_{\bm \sigma} \hat{U}_{\bm{s \sigma}} c_{\bm \sigma},}
where now the precomputed transformation matrices are 
\EQ{\hat{U}_{\bm{s\sigma}} = \phi_\alpha (\hat{x}_i) \phi_\beta (\hat{x}_j) \phi_\gamma (\hat{x}_k) .} 

\item[Step 4:] We form the appropriate integrand of $dB_{n}$, 
\EQ{b(\hat\x_{\bm s})  = \hat{w}_{\bm s} e^{3\hat\x^2_{\bm s}/2} \cf(\hat\x_{\bm s}) f(\x_{\bm{s}}) .}  

\item[Step 4:] The inverse transform of $b(\hat\x_{\bm s}) $ completes the required integration
\EQ{dB_{\bm\sigma} = \sum_{\bm s} \hat{U}_{\bm s \sigma}^* b(\hat\x_{\bm s}) .}
\end{itemize}

%%%%%%%%%%%%%%%%%%%%%%%%%%%%%%%%%%%%%%%%%%%%%%%%%%%

\section{Accuracy of algorithm \label{accuracy}}
In this section, we discuss how we can control the accuracy of our algorithm, and quantify the accuracy of our approach of both spatial and temporal integration using various measures.

\subsection{$k$-space quadrature grid \label{kspace_grids}}
Both the deterministic and noise terms involve $k$-space integrals which are evaluated as approximate quadrature sums [evaluating \eeref{Phielement}, and \eeref{fnoise} respectively].  To control the accuracy of these steps, we can vary the size of the appropriate $k$-space quadrature grid.  

For a given $\ecut$, the transform to momentum space is exactly invertible $[{\bf j} (\x) \xrightarrow{W} {\bf j} (\kk) \xrightarrow{W^\dagger} {\bf j} (\x)]$ if the quadrature grid sizes are chosen by $N_x \geq N_x^0$, $N_k \geq N_k^0$, where $N_x^0 \equiv 2M_x-1, N_k^0 \equiv 2M_x$.  We choose an even number of $k$-grid points to avoid a quadrature point at $\kk = 0$ where both the deterministic and noise terms are singular.  To investigate the effect of the $k$-grid on the accuracy of our method, we vary the number of $k$ quadrature points used by 
\eqn{
N_k &=& N_k^0+\Delta N_k, \label{kquadgridcheck_pot}\\
N_k^\prime &=& N_k^{0\prime}+\Delta N_k^\prime,\label{kquadgridcheck_noise}}
where $N_k$ is the total number of quadrature grid points, and $\Delta N_k$ is the number of quadrature points added to the reference value $N_k^0$.  \eeref{kquadgridcheck_pot} refers to the number of quadrature points used to evaluate the deterministic term, i.e. the quadrature points are based on the weight function $e^{-2 k_i^2}$ [see \eeref{Tquad}].  \eeref{kquadgridcheck_noise} refers to the quadrature grid for evaluating the noise term, i.e. the quadrature points are based on the weight function $e^{- k_i^2}$ [see \eeref{Upoly}].  For the remainder of this section we associate $\Delta N_k$ and $\Delta N_k^\prime$ with the quadrature grid for the deterministic and noise terms respectively.

\subsection{Effective potential term convergence for a breathing mode}
\begin{figure}[!t]
\begin{center}
\includegraphics[width=\columnwidth]{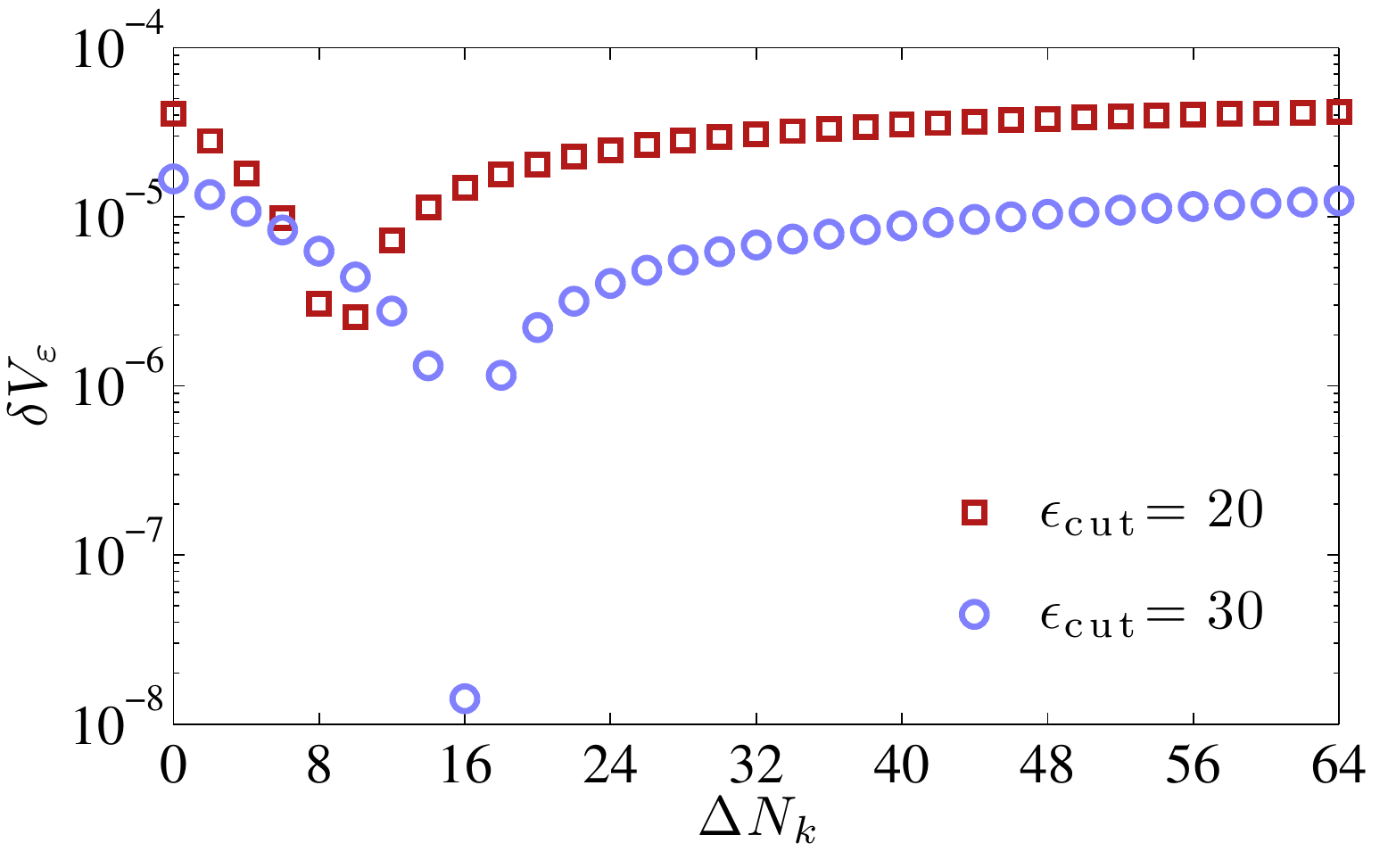}
\caption{(Color online) Relative error in the potential $V_\ve(\x)$ with varying $\Delta N_k$ [see \eref{delta_Ve}], for a Gaussian wave function with $N = 1000, \sigma = 1, \kappa = 0.5$.   $\ecut = 20$ $\ecut = 30$. The two curves correspond to $\ecut = 20$ (red squares), and $\ecut = 30$ (blue circles).}
\label{fig:errorVe}
\end{center}
\end{figure}

A Gaussian wave function undergoing radial breathing provides a rare example where we can analytically calculate $V_\ve (\x)$ for a case where $\nabla\cdot{\bf j} \neq 0$.  Here we consider the Gaussian wave function 
\eq{Gausswf}
{\cf(\x) = \frac{\sqrt{N}}{(\pi \sigma^2)^{3/4}} e^{-r^2/2\sigma^2 + i \kappa r^2/2}, }
which corresponds to a radial breathing oscillation. This wave function has a non-zero current density, given by
\eq{JGauss}
{ {\bf j} (\x) = \kappa \x |\psi |^2,}
from which we find [using \eeref{Ve}] the radially symmetric potential 
\eq{VGauss}
{V^A_\ve(r) = -\frac{N^2 \mathcal{M} \kappa \sigma}{\sqrt{2} (\pi \sigma^2)^{3/4}} g(r/\sigma),} 
where
\eq{g_for_V}
{ g(x) = \left[ \sqrt{\frac{2}{\pi}} +  \left(\frac{1}{\sqrt{2}x}  - \sqrt{2}x\right) e^{-x^2} {\rm erfi} (x) \right], }
with ${\rm erfi} (x ) \equiv -i {\rm erf} (ix)$.  The potential is most significant at $r = 0$, where 
\eq{Ve_Gauss_zero}
{V^A_\ve (r=0) = - \frac{N^2 \mathcal{M} \kappa \sigma}{\sqrt{2}(\pi \sigma^2)^{3/4}} \sqrt{\frac{8}{\pi}}.}
To quantify the accuracy of our numerical evaluation of $V_\ve(\x)$, we use the relative error measure 
\eq{delta_Ve}
{\delta V_\ve = 1 - \left| \frac{V_\ve(r=0)}{V^A_\ve (r=0)} \right|.}
In \fref{fig:errorVe} we plot $\delta V_\ve$ as a function of $\Delta N_k$, for the Gaussian wave function [\eeref{Gausswf}], with $N = 1000, \sigma  = 1,$ and $\kappa = 0.5$.  We show the relative error for $V_\ve$ evaluated with $\ecut = 20$ and $\ecut = 30$, corresponding to $M_x = 19$ and $M_x = 29$ respectively.  In both cases $\delta V_\ve$ rapidly converges with $\Delta N_k$.  We see that in general our algorithm does not become increasingly accurate with $\Delta N_k$, with no gain in accuracy for $\Delta N_k > 16$.  However in general we see good accuracy, with $\delta V_\ve < 10^{-4}$ for all $\Delta N_k$.

%====================================================
%====================================================

\subsection{Matrix elements convergence}
Here we test the accuracy of calculating the matrix elements [Eqs.~\eref{Gn}-\eref{Bnm} ] for a randomized state $c_{\alpha \beta \gamma}$.  A high energy randomized state will test the accuracy of both low energy modes and modes near the cutoff, thus will be a useful test for the suitability of our algorithm to finite temperature non-equilibrium dynamics.  

We use a random state of the form
\eq{randinit}
{c_{\bm\sigma} = \eta_{\bm\sigma} + i \xi_{\bm\sigma},}
where $\{\eta_{\bm\sigma}\}$ and $\{\xi_{\bm\sigma}\}$, are normally distributed Gaussian  random variables with zero mean and unit variance. For the results in this section, we use a cutoff of $\ecut = 20$ so $M_x = 19$.  We renormalize our initial field so that $N = 1\times10^4$, and use scattering reservoir parameters of $\mathcal{M} = 0.005, T = 5$. 

\subsubsection{Effective potential term matrix elements}

We first consider $S_{\bm\sigma}$, the scattering potential matrix elements given by \eeref{Sn}.  To test the accuracy of our method we calculate the relative error of the matrix elements, given by
\eq{Serror}
{\mathcal{E}_S (\Delta N_k) = \frac{ |S_{\bm\sigma} (\Delta N_k) -S_{\bm\sigma}^{A} |}{|S_{\bm\sigma}^{A}|},} 
where $S_{\bm\sigma}(\Delta N_k)$ is the scattering matrix element calculated with a $k$-space quadrature grid specified by $\Delta N_k$, and $S_{\bm\sigma}^{A}$ is a more accurate matrix element calculation.  We calculate ${S_{\bm\sigma}^{A}}$ using a $k$-space quadrature grid with $\Delta N_k = 128$ quadrature points.  

\begin{figure}[!t]
\begin{center}
\includegraphics[width=\columnwidth]{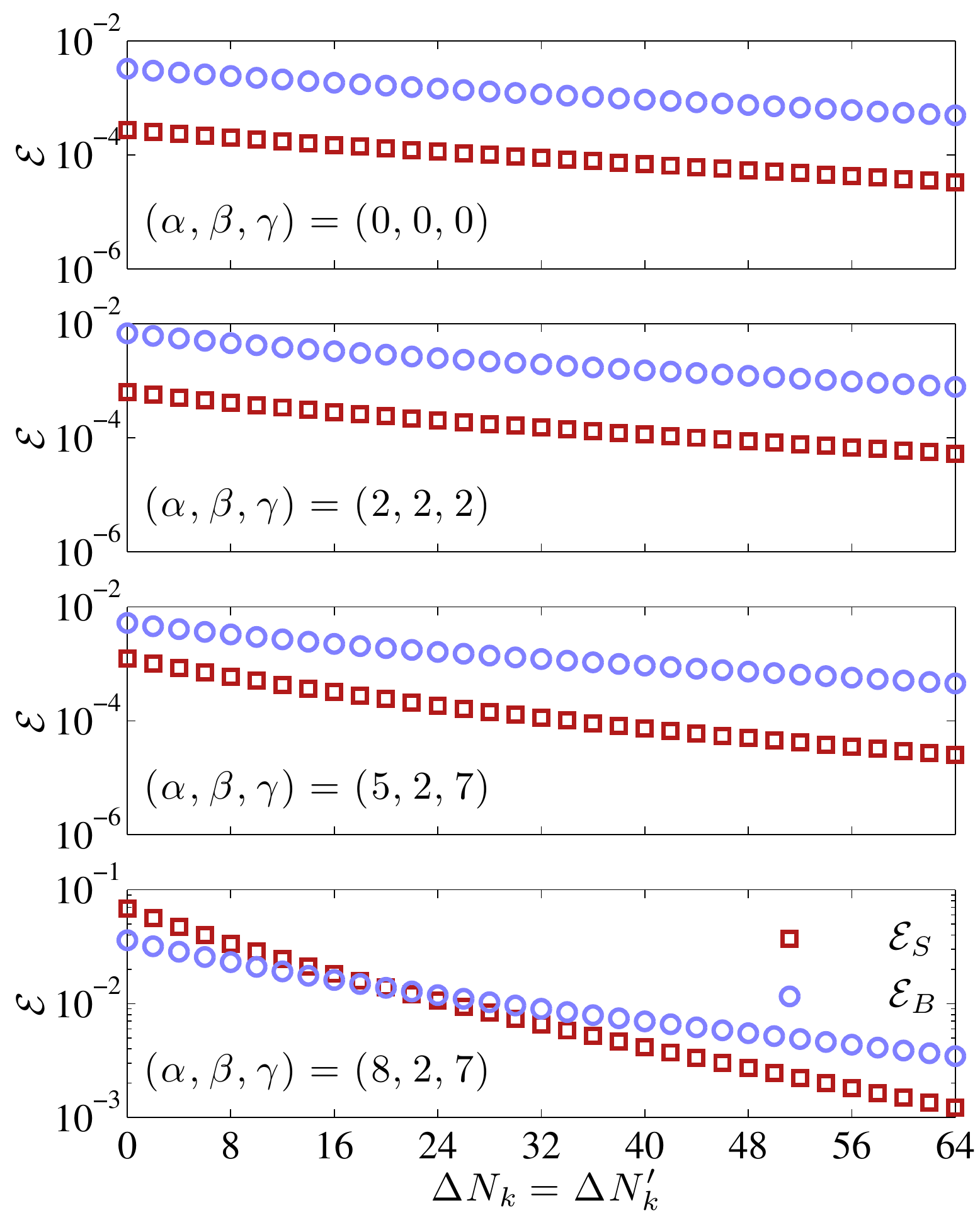}
\caption{(Color online) Relative error in individual matrix elements as $\Delta N_k$ (or $\Delta N_k^\prime$) is varied.  The error in the scattering potential matrix elements $\mathcal{E}_S(\Delta N_k)$ (red squares), and scattering noise matrix elements $\mathcal{E}_B(\Delta N_k^\prime)$ (blue circles), are shown for various single-particle basis states.  All cases correspond to $\ecut = 20$.}
\label{fig:SnBnconv}
\end{center}
\end{figure}

In \fref{fig:SnBnconv} we show $\mathcal{E}_S(\Delta N_k)$ for some representative cases of ${\bm\sigma}=(\alpha,\beta,\gamma)$.  We see $\mathcal{E}_S$ is smaller for lower-order matrix elements.  In general, $\mathcal{E}_S$ decreases with $\Delta N_k$ for all basis coefficients.  

%%====================================================
%====================================================

\subsubsection{Noise matrix elements}
We now test the accuracy of the scattering noise matrix elements $dB_{\bm\sigma}$ [see \eeref{Bnm}].  Here we quantify the error with the measure
\eq{Berror}
{\mathcal{E}_B (\Delta N_k^\prime) = \frac{ |dB_{\bm\sigma}(\Delta N_k^\prime) -dB_{\bm\sigma}^{A} |}{|dB_{\bm\sigma}^{A}|},} 
where $dB_{\bm\sigma}({\Delta N_k^\prime})$ are the scattering noise matrix elements calculated with a $k$-space quadrature grid specified by $\Delta N_k^\prime$, where the quadrature weight function is that appropriate to the noise term.  Here the reference ${dB_{\bm\sigma}^{A}}$ is calculated using a $k$-space quadrature grid with $\Delta N_k^\prime = 128$ points.  

In \fref{fig:SnBnconv} we plot $\mathcal{E}_B(\Delta N_k^\prime)$ for various ${\bm\sigma}=(\alpha,\beta,\gamma)$, finding a smaller error with increasing grid size. The noise term is generally less accurate than the deterministic term, with $\mathcal{E}_B$ over an order of magnitude larger than $\mathcal{E}_S$ for low-order matrix elements.  For higher-order matrix elements, it becomes hard to distinguish between the accuracy of the noise and deterministic term, with both having a similar error.

%%====================================================
\begin{figure}[!t]
\begin{center}
\includegraphics[width=\columnwidth]{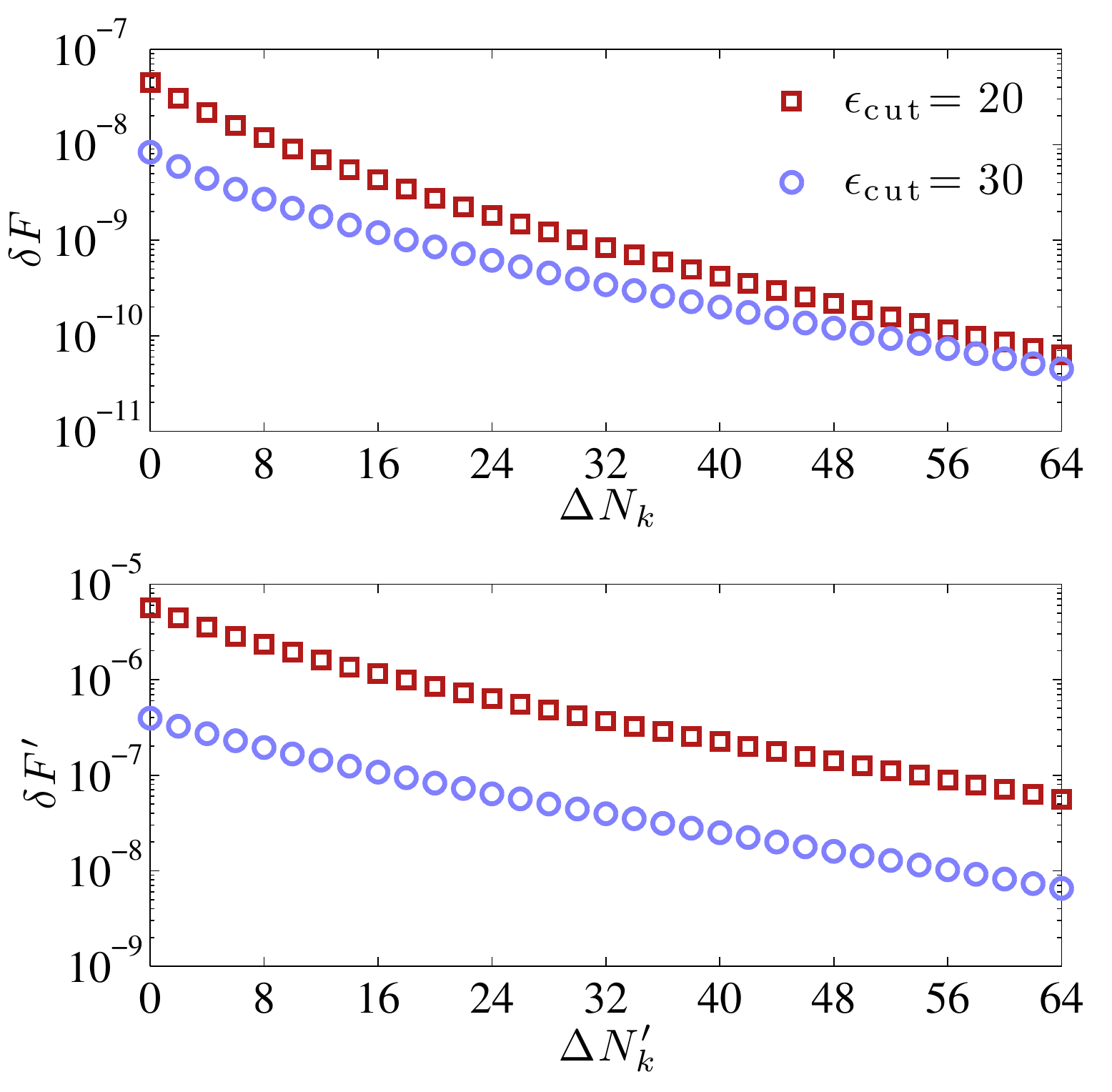}
\caption{(Color online) Relative error in the combined matrix elements [see \eeref{deltaF}], for $\ecut = 20$ (red squares) and $\ecut = 30$ (blue circles).  Top: $\delta F(\Delta N_k)$, showing the effect of the $k$-space quadrature grid corresponding to the potential term, with $\Delta N_k^\prime = 0$.  Bottom: $\delta F^\prime(\Delta N_k^\prime)$, showing the effect of the $k$-space quadrature grid corresponding to the noise term, with $\Delta N_k = 0$.}
\label{fig:Fnconv}
\end{center}
\end{figure}
%====================================================

\subsubsection{All nonlinear matrix elements}
Finally we consider the combination of all nonlinear matrix elements, $F_{\bm\sigma} = G_{\bm\sigma}+S_{\bm\sigma}+dB_{\bm\sigma}$. To evaluate $G_{\bm\sigma}$, we use a nonlinearity constant of $C_{\rm NL} = 0.005$.  We use the relative error measure 
\eq{deltaF}
{\delta F \equiv \frac{||F_{\bm\sigma} - F_{\bm\sigma}^A||^2}{||F_{\bm\sigma}^A||^2} ,}
where 
\eq{Lambdameasure}{
||\Lambda_{\bm\sigma} ||  \equiv \sum_{\bm\sigma} |\Lambda_{\bm\sigma}|^2,}
where $F_{\bm\sigma}$ are the approximate matrix elements calculated by our algorithm, and $F_{\bm\sigma}^A$ are more accurately calculated matrix elements.  $\delta F$ includes both the deterministic and noise terms from the scattering processes, so we can measure the combined effect of these terms on the accuracy of our algorithm.  

Since both the deterministic and noise terms have independent quadrature grids which affect the accuracy, we use two different measures of $\delta F$.  We refer to $\delta F$ when we determine $F_{\bm{\sigma}}^A$ with a quadrature grid of $\Delta N_k = 128$ for the deterministic term, and $\Delta N_k^\prime = 0$ for the noise term.   Secondly, we refer to $\delta F^\prime$ when we determine $F_{\bm{\sigma}}^A$ with a quadrature grid of $\Delta N_k = 0$ for the deterministic term, and $\Delta N_k^\prime =128$ for the noise term.  These two different measures allow us to examine the effect each $k$-space quadrature grid we employ has on the accuracy of the combined matrix elements.  

In \fref{fig:Fnconv} we show results for $\delta F$ with varying $\Delta N_k$, and $\delta F^\prime$ with varying $\Delta N_k^\prime$. We see excellent accuracy for all $\Delta N_k$ in both measures, with the relative error reducing with increasing grid size in either case.  Since $\delta F$ is weighted by the size of each matrix element, a large relative error in small matrix elements has only a small effect on $\delta F$. Thus the larger error seen in high energy matrix elements in \fref{fig:SnBnconv} has a minimal effect on $\delta F$.  For larger $\ecut$ we see a smaller error, with little change in the rate of convergence. 

\subsection{Propagation convergence \label{sec:propconv}}
In this section we present evolution convergence results for our algorithm for the scattering SPGPE.  To test the evolution of our algorithm, we propagate an initial random state given by \eeref{randinit}, with an energy cutoff of $\ecut = 20$,   and we normalize our c-field to $N(t=0) = 1\times10^4$ $^{87}{\rm Rb}$ atoms, so the nonlinearity constant is $C = 0.02$.  We choose reservoir parameters of $T = 20, \mathcal{M} = 0.005$, and evolve our initial state with the scattering SPGPE using the semi-implicit Euler algorithm for one trap cycle with a final time of $\tau = 2\pi$.  

In the following sections, we will test the accuracy of our evolution by examining the dependence on time step size $\Delta t$, and on both $\Delta N_k$ and $\Delta N_k^\prime$.  As the SPGPE is a stochastic equation of motion, we look at ensemble averages of 500 trajectories for each parameter $\{\Delta t, \Delta N_k, \Delta N_k^\prime\}$ considered.  We consider a unique initial condition for each trajectory.  

%%====================================================
\begin{figure}[!t]
\begin{center}
\includegraphics[width=\columnwidth]{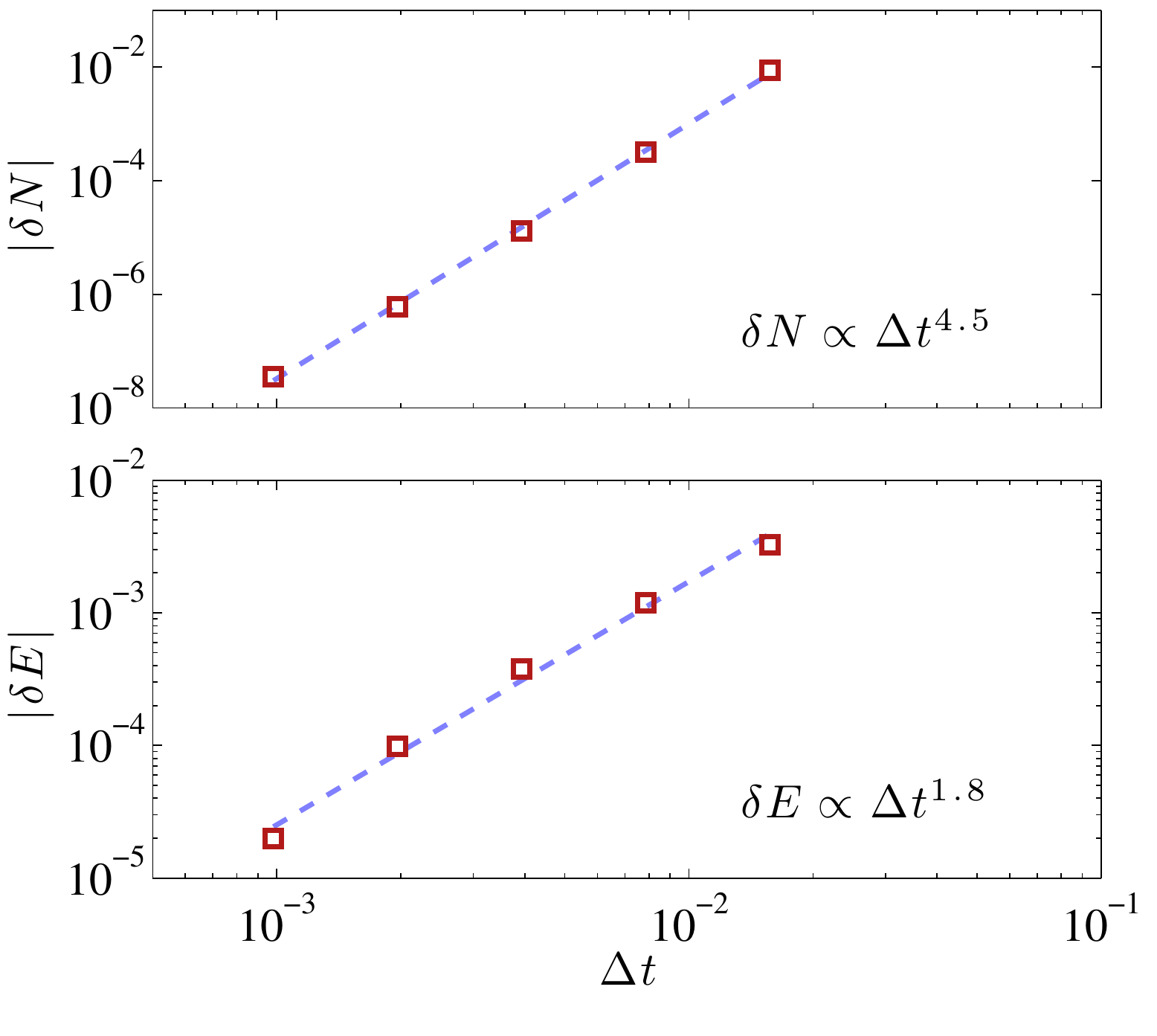}
\caption{(Color online) Number ($\delta N$) and energy ($\delta E$) convergence with step size $\Delta t$.  The relative error is determined after $t=  2\pi$, where $\Delta t$ corresponds to 6400, 3200, 1600, 800, and 400 time steps in order of increasing $\Delta t$.  Simulation results (red squares) are fitted by the labeled power laws.  }
\label{fig:NE_tconv}
\end{center}
\end{figure}
%====================================================

\subsubsection{Time step convergence}
In this section we consider the case $\Delta N_k = \Delta N_k^\prime= 0$, and examine the convergence with $\Delta t$.  To quantify our accuracy, we use the following measures:
\eq{deltaN}
{ \delta N  = \left\langle \frac{N(t=0) -  \sum_{\bm\sigma} |c_{\bm\sigma}(\tau)|^2}{N(t=0)} \right\rangle ,} 
\eq{deltaE}
{\delta E  = \left\langle \frac{E (\tau) - E^A(\tau) }{E^A (\tau)} \right\rangle, } 
\eq{deltac}
{ \delta c_{\bm\sigma} = \left\langle \frac{|c_{\bm\sigma} (\tau) - c_{\bm\sigma}^A(\tau) |^2}{| c_{\bm\sigma}^A(\tau) |^2} \right\rangle ^{1/2},} 
\eq{deltaX}
{ \delta X  = \left\langle\frac{ \sum_{\bm\sigma} |c_{\bm\sigma} (\tau) - c_{\bm\sigma}^A(\tau) |^2}{ \sum_{\bm\sigma}| c_{\bm\sigma}^A(\tau) |^2} \right\rangle ^{1/2}.}
where $\delta N$ is the change in normalization, $\delta E$ is the relative change in energy, where the c-field energy is given by
\eq{energy}
{E = \intV{\x} \cf^* H_{\rm sp} \cf + \frac{C}{2} \intV{\x} |\cf|^4 .}
$\delta c_{\bm\sigma}$ is the relative change in individual mode amplitudes, and $\delta X$ is the relative difference of all mode amplitudes.  Quantities with a subscript `A' denote that quantity is calculated from a more accurate simulation, and angle brackets represent an ensemble average.  Since the scattering SPGPE is formally number conserving, $\delta N$ provides an immediate indication of accuracy without needing a more accurate simulation. 

%%====================================================
\begin{figure}[!t]
\begin{center}
\includegraphics[width=\columnwidth]{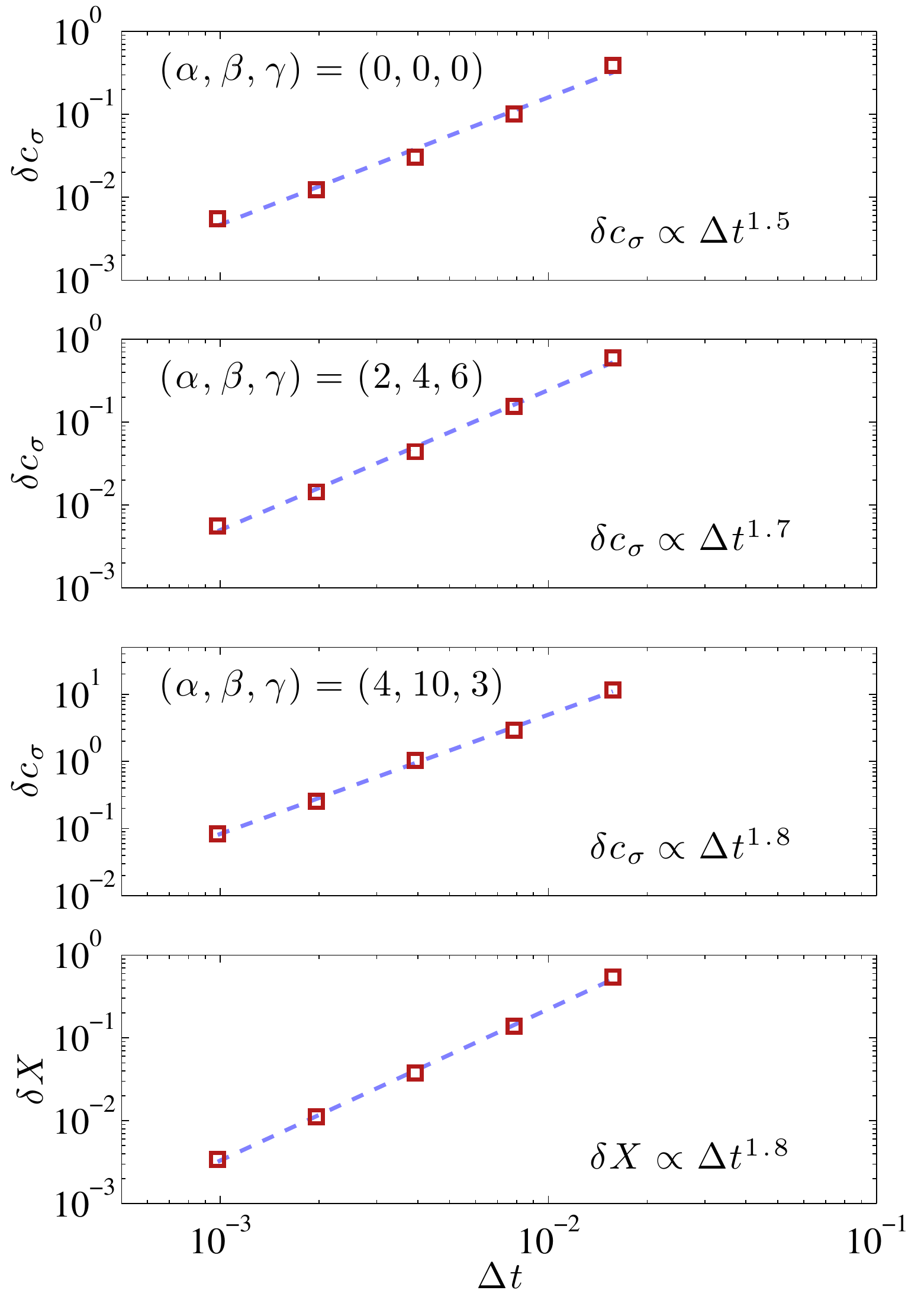}
\caption{(Color online) Mode amplitude convergence with $\Delta t$. $\delta c_{\bm\sigma}$ represent the error in individual modes [see \eeref{deltac}].  $\delta X$ represents the error over all \rC-region modes [see \eeref{deltaX}].Simulation results (red squares) are fitted by the labeled power laws. }
\label{fig:cf_tconv}
\end{center}
\end{figure}
%====================================================

To test the convergence with $\Delta t$, we perform 500 simulations with a time step of $\Delta t^A = \tau / 12800$, i.e. use 12800 steps in the evolution time of 1 trap period ($\tau = 2\pi$).  Simulations with a step size $\Delta t^A$ are our accurate simulations for considering the convergence tests.  Each simulation with step size $\Delta t^A$ has a different random initial condition.  To understand the effects of $\Delta t$ on convergence we repeat the SPGPE simulation for each trajectory using an identical initial state, while reducing the time step size to $\delta t = \Delta t^A / 2^p$, for $p = \{1,2,3,4,5\}$.  To test stochastic convergence for each trajectory, we use the identical noise as for the accurate simulation.  

The first two measures of numerical accuracy [\eref{deltaN} and \eref{deltaE}] represent measures of weak convergence \cite{SM}.  These two measures show for an ensemble of simulations at a given time step $\Delta t$, the difference between the average number and energy, and their more accurate respective value.  \fref{fig:NE_tconv} shows $\delta N$ and $\delta E$, where we observe good convergence with $\Delta t$.  The relative errors are at the level of less than 1\% for all time steps considered. 

 %%====================================================
\begin{figure}[!t]
\begin{center}
\includegraphics[width=\columnwidth]{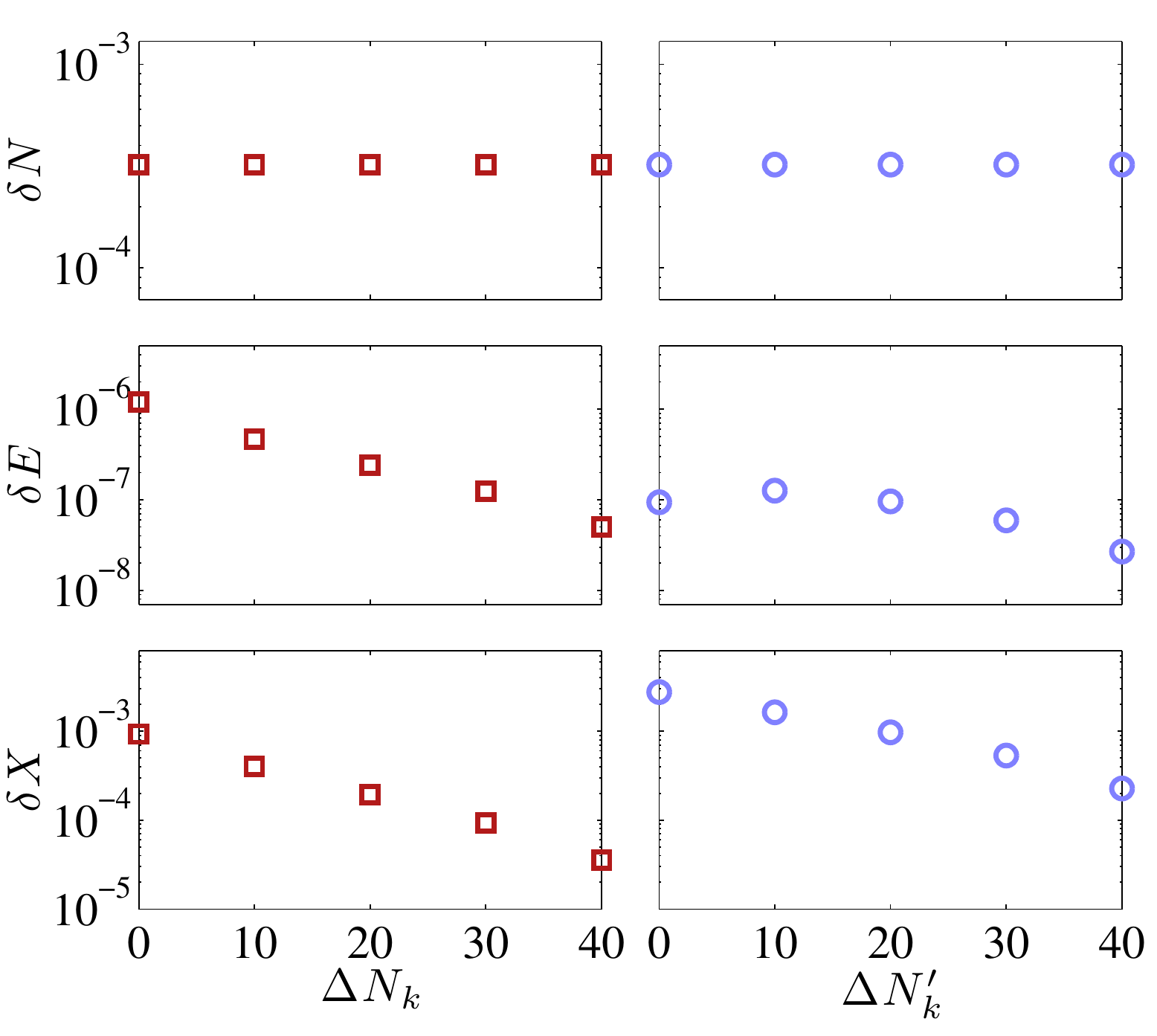}
\caption{(Color online) Measures of convergence with $k$-space quadrature grid size, for simulations with fixed $\Delta t = 0.001$.  Red squares: Accuracy with varying $\delta N_k$ for fixed $\delta N_k^\prime = 0$, where the accurate simulations have $\Delta N_k = 50$.  Blue circles: Accuracy with varying $\delta N_k^\prime$ for fixed $\delta N_k = 0$, where the accurate simulations have $\Delta N_k^\prime = 50$}
\label{fig:dNkconv}
\end{center}
\end{figure}
%====================================================
 
The final two measures of error [\eref{deltac} and \eref{deltaX}] show the root mean squared difference of the mode amplitudes from our simulations with an accurate value.  As these are the solutions to the scattering SPGPE, these measures determine stochastic convergence in the strong sense \cite{SM}.  $\delta c_{\bm\sigma}$ shows the relative error in the individual mode amplitudes.  In \fref{fig:cf_tconv} we show $\delta c_{\bm\sigma}$ for the basis-states $\bm\sigma  = (0,0,0), (2,4,6)$, and $(4,10,3)$.  In all cases we see convergence faster than $\Delta t^{1.5}$, as we reduce $\Delta t$.  The error increases for higher-order modes with quite large relative error in the $(4,10,3)$ state, which is near the cutoff.  To give an idea of the effect all modes have on the accuracy of a simulation we calculate $\delta X$, which takes into account the relative error of each mode.  In \fref{fig:cf_tconv} we see $\delta X \propto \Delta t^{1.8}$, with accuracy of greater greater than 1\% for the smaller time steps.  We see that the larger error in $\delta c_{\bm\sigma}$ for high energy modes has little effect on $\delta X$, since $\delta X$ depends on the size of each element.  

Figs \ref{fig:NE_tconv} and \ref{fig:cf_tconv} show that all measures of accuracy converge at least as fast as the strong vector semi-implicit Euler method of Ref.~\cite{SM}, which converges as $\Delta t^1$ with decreasing $\Delta t$ for both strong and weak measures of convergence.  In principle it is possible to generate higher order algorithms, but this task becomes rapidly complex with increasing desired accuracy, requiring the sampling of many additional auxiliary noise terms~\cite{SM}.  Here we find quite fast convergence using the weak semi-implicit Euler method, suggesting that the noise is commutative, in which case the strong and weak methods are equivalent. 

\subsubsection{ $k$-space grid convergence}

In this section we investigate the effect that the $k$-space quadrature grid has on the convergence of our algorithm.  We consider the measures $\delta N, \delta E$, and $\delta X$, for fixed $\Delta t = 0.001$, i.e. 800 integration steps.  For each trajectory (as described above), we evolve the scattering SPGPE with the same initial conditions and same noise, while varying $\Delta N_k$ and $\Delta N_k^\prime$ independently.  

Firstly we consider the case of varying $\Delta N_k$ with fixed $\Delta N_k^\prime = 0$, where we plot $\delta N, \delta E$, and $\delta X$ calculated after $\tau = 2\pi$ in \fref{fig:dNkconv} (red squares).  To calculate these measures of accuracy, we use an accurate value based on simulations with $\Delta N_k = 50$.  Varying $\Delta N_k$ has no effect on $\delta N$, while $\delta E$ and $\delta X$ both converge rapidly.  

The case of  varying $\Delta N_k^\prime$ with fixed $\Delta N_k = 0$ is shown in \fref{fig:dNkconv} (blue circles), where we see $\delta N, \delta E$, and $\delta X$  behave in a similar manner.  In all cases for either varying $\Delta N_k$ or $\Delta N_k^\prime$, the accuracy is very good even for the cases $\Delta N_k = 0$ and $\delta N_k^\prime=0$, consistent with \fref{fig:Fnconv}.

The good accuracy for all $k$-space quadrature grid sizes shown in \fref{fig:dNkconv} means that increasing $\Delta N_k$ or $\Delta N_k^\prime$ has only a minor benefit on the accuracy of our algorithm for calculating the SPGPE matrix elements.  The most significant gains arise from controlling the time step size $\Delta t$.
 
%====================================================
\begin{figure}[!t]
\begin{center}
\includegraphics[width=\columnwidth]{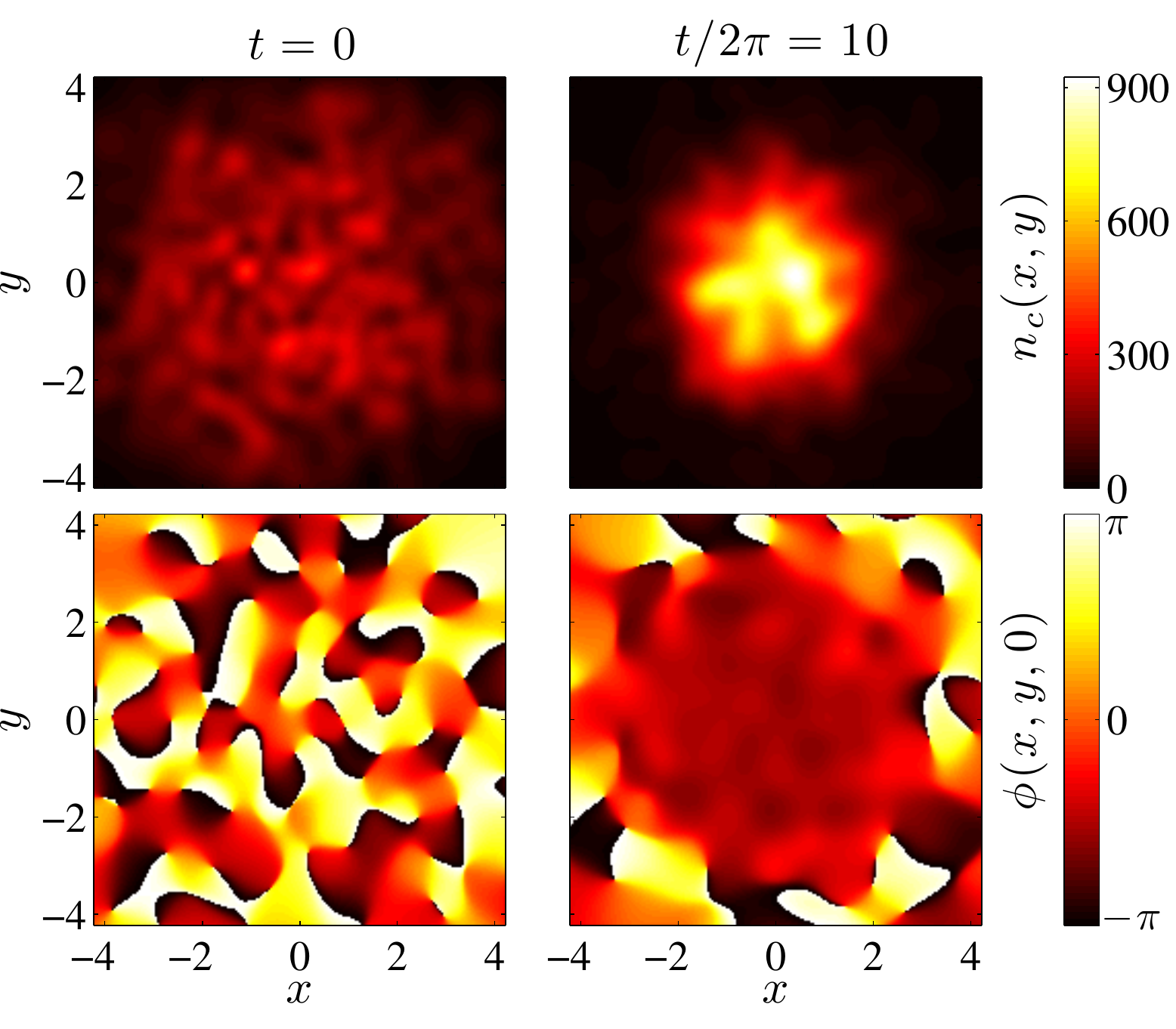}
\caption{(Color online) Column densities and phase slices (through the $z = 0$ plane), for the SPGPE evolution of an initial random state with $T = 20$, $\mathcal{M} = 0.01$.  The initial random state at $t = 0$  evolves into equilibrium, where we show the final state after 10 trap cycles of evolution.  }
\label{fig:density}
\end{center}
\end{figure}
%====================================================

\subsection{Random state evolution to equilibrium}
Having characterized the accuracy of our algorithm, we now look at a practical example of using the SPGPE.  Here we evolve a random initial state into thermal equilibrium with the SPGPE, and examine some thermodynamic quantities which are commonly of interest in c-field calculations. 

As in Sec.~\ref{sec:propconv} we use a randomized initial state for each trajectory given by \eeref{randinit}, with normalization $1\times10^4$ $^{87}{\rm Rb}$ atoms, and use an energy cutoff of $\ecut = 20$.  We evolve the scattering SPGPE for 10 trap cycles using $1600$ integration steps per trap cycle, and set $\Delta N_k = \Delta N_k^\prime = 0$.  To visualize our simulations, we look at the column density given by
\eq{coldens}
{n_c(x,y) = \int dz |\cf(\x)|^2.}
In \fref{fig:density} we show the column density and phase (slice through $z = 0$) of our initial random state.  At $t/2\pi = 10$ once equilibrium has been reached, we see phase coherence has developed over regions of finite c-field density.  Fluctuations in the density are caused by both the scattering noise and from fluctuations arising from PGPE evolution, and are characteristic of a finite temperature equilibrium state.  

%====================================================
\begin{figure}[!t]
\begin{center}
\includegraphics[width=\columnwidth]{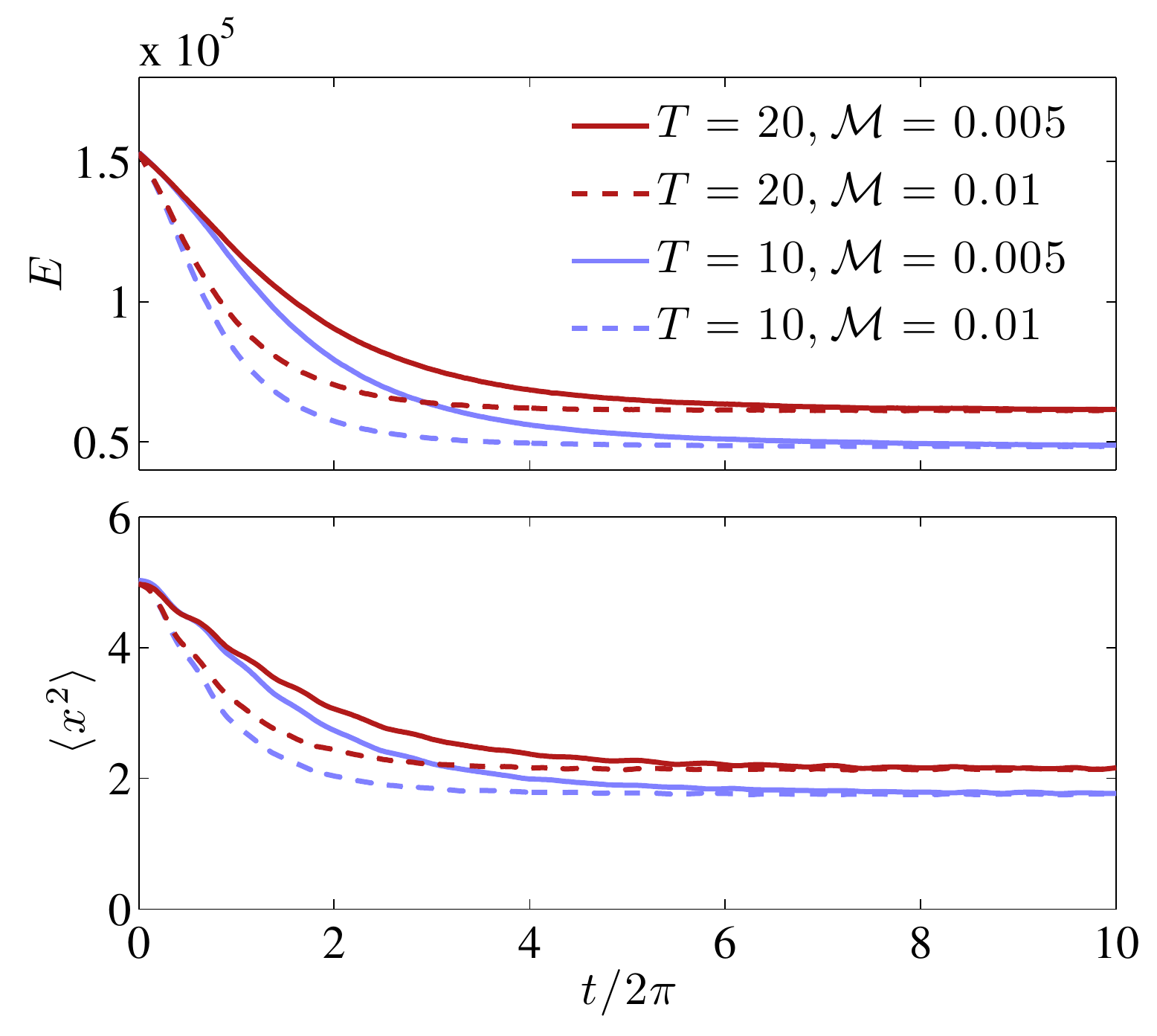}
\caption{(Color online) Energy ($E$) and system width ($\langle x^2 \rangle$) averaged over 100 trajectories, showing the evolution of a random initial state to thermal equilibrium.  (red curves) SPGPE simulations for $T = 20$, with $\mathcal{M} = 0.005$ (dashed) and $\mathcal{M} = 0.01$ (solid).  (blue curves) SPGPE simulations for $T = 10$, with $\mathcal{M} = 0.005$ (dashed) and $\mathcal{M} = 0.01$ (solid).   }
\label{fig:rand2equilb}
\end{center}
\end{figure}
%====================================================

The images in \fref{fig:density} highlight that individual trajectories of the SPGPE can be thought of as showing the corresponding dynamics of a single experimental run \cite{Blakie08a}.  However when calculating thermodynamics quantities or correlation functions, we must calculate ensemble averages.  \fref{fig:rand2equilb} shows the c-field energy ($E$), and system width ($\langle x^2\rangle$), where we calculate these quantities as an ensemble average over 100 trajectories of scattering SPGPE evolution.  Results are shown for two different reservoir temperatures of $T = 10$ and $T = 20$. For each temperature we consider reservoir interaction amplitudes of $\mathcal{M} = 0.005$ and $\mathcal{M} = 0.01$.  We see the high energy of the random initial state rapidly decreases as the system evolves towards equilibrium.  Note the decay time decreases with larger $\mathcal{M}$, and the final equilibrium energy increases with $T$.  The evolution of the system width $\langle x^2 \rangle$ follows a similar trend to the energy.  The large spread of density in the initial state [see \fref{fig:density}] decreases as a well defined Bose-Einstein condensate emerges as equilibrium is attained.  The equilibrium system width increases with temperature, due to an increased thermal density contained in the c-field.  The results in \fref{fig:rand2equilb} show that for a given temperature, the equilibrium reached is independent of $\mathcal{M}$.  This is an important result to verify, as it gives a good test of the implementation of the noise \cite{Rooney12a}. 

\section{Conclusions}
We have presented a method to numerically solve the SPGPE for a finite temperature Bose gas in a three-dimensional harmonic trap.  Our algorithm allows us to accurately and efficiently implement the terms for the scattering reservoir interaction, while maintaining a consistent implementation of the projector via a spectral approach. 

We have extensively tested the accuracy of our evaluation of matrix elements associated the deterministic and noise terms, steps which are not exact within our Gauss-Hermite integration scheme.  We have shown how the accuracy of these terms can be controlled with increasing order of $k$-space quadrature grid.  While this leads to increased accuracy, for the systems considered in this paper our procedure showed good accuracy for all grid sizes considered.  However care should be taken when implementing this algorithm in any novel system, to ensure similar convergence of matrix elements is achieved.  

We presented convergence rates with respect to time-step size used in our weak semi-implicit Euler implementation. For all measures considered the weak semi-implicit Euler method exhibits rapid convergence (faster than $\Delta t^1$), in both the strong and weak sense, although the precise rate depends on the observable in question. The algorithm converges more rapidly than the strong semi-implicit algorithm, indicating that the scattering noise may be commutative~\cite{SM}. We found that the choice of time step is more important to the accuracy of SPGPE evolution than the $k$-space quadrature grid size.  

Finally we demonstrated typical usage of the SPGPE by examining the evolution of a random initial state to thermal equilibrium. The thermodynamic equilibrium results are found to to be independent of $\mathcal{M}$, as must hold for a physically consistent implementation of the reservoir interaction. 

\section*{Acknowledgments}
We acknowledge the use of the University of Otago Vulcan cluster, and the Victoria University of Wellington Sci-Fac HPC facility.  SJR thanks Victoria University of Wellington where this work commenced for their hospitality. This work was supported by the University of Otago (SJR),  the Marsden Fund of New Zealand (PBB, ASB), and a Rutherford Discovery Fellowship, administered by the Royal Society of New Zealand (ASB).

%

%******************************************************************************
\end{document}